\documentclass[preprint,twocolumn]{aastex631} 

\usepackage[printonlyused]{acronym} 
\usepackage{float}
\usepackage[outline]{contour} 

\newcommand\blfootnote[1]{%
  \begingroup
  \renewcommand\thefootnote{}\footnote{#1}%
  \addtocounter{footnote}{-1}%
  \endgroup
}

\newcounter{obsnotelabel}
\newcommand{\obsnote}[1]{\refstepcounter{obsnotelabel}\label{#1}}
\newcommand{\objname}{282P}
\newcommand{\objnameFull}{282P/(323137)~2003~BM$_{80}$} 

\shorttitle{282P: An Active Quasi-Hilda Object}
\shortauthors{Chandler, Oldroyd \& Trujillo}

\usepackage{graphicx}
\usepackage[percent]{overpic} 

\begin{document}


\title{Migratory Outbursting Quasi-Hilda Object 282P/(323137) 2003 BM80}

\correspondingauthor{Colin Orion Chandler}
\email{orion@nau.edu}

\author[0000-0001-7335-1715]{Colin Orion Chandler}
\affiliation{Department of Astronomy and Planetary Science, Northern Arizona University, PO Box 6010, Flagstaff, AZ 86011, USA}

\author[0000-0001-5750-4953]{William J. Oldroyd}
\affiliation{Department of Astronomy and Planetary Science, Northern Arizona University, PO Box 6010, Flagstaff, AZ 86011, USA}

\author[0000-0001-9859-0894]{Chadwick A. Trujillo}
\affiliation{Department of Astronomy and Planetary Science, Northern Arizona University, PO Box 6010, Flagstaff, AZ 86011, USA}



\begin{abstract}
\label{282P:Abstract}
We report object \objnameFull{} is undergoing a sustained activity outburst, lasting over 15 months thus far. These findings stem in part from our \acs{NASA} Partner Citizen Science project \textit{Active Asteroids} (\url{http://activeasteroids.net}), which we introduce here. 
We acquired new observations of \objname{} via our observing campaign (\acf{VATT}, \acf{LDT}, and the Gemini South telescope), confirming \objname{} was active on UT 2022 June 7, some 15 months after 2021 March images showed activity in the 2021--2022 epoch. 
We classify \objname{} as a member of the \acp{QHO}, a group of dynamically unstable objects found in an orbital region similar to, but distinct in their dynamical characteristics to, the Hilda asteroids (objects in 3:2 resonance with Jupiter). Our dynamical simulations show \objname{} has undergone at least five close encounters with Jupiter and one with Saturn over the last 180 years. \objname{} was most likely a Centaur or \ac{JFC} 250 years ago. In 350 years, following some 15 strong Jovian interactions, \objname{} will most likely migrate to become a \ac{JFC} or, less likely, an \acl{OMBA} orbit. These migrations highlight a dynamical pathway connecting Centaurs and \acp{JFC} with Quasi-Hildas and, potentially, active asteroids. Synthesizing these results with our thermodynamical modeling and new activity observations, we find volatile sublimation is the primary activity mechanism. Observations of a quiescent \objname{}, which we anticipate will be possible in 2023, will help confirm our hypothesis by measuring a rotation period and ascertaining spectral type. 
\end{abstract}

\keywords{minor planets, Quasi-Hildas: individual (282P), comets: individual (282P)}

\blfootnote{Based on observations obtained at the international Gemini Observatory, a program of NSF’s NOIRLab, which is managed by the Association of Universities for Research in Astronomy (AURA) under a cooperative agreement with the National Science Foundation on behalf of the Gemini Observatory partnership: the National Science Foundation (United States), National Research Council (Canada), Agencia Nacional de Investigaci\'{o}n y Desarrollo (Chile), Ministerio de Ciencia, Tecnolog\'{i}a e Innovaci\'{o}n (Argentina), Minist\'{e}rio da Ci\^{e}ncia, Tecnologia, Inova\c{c}\~{o}es e Comunica\c{c}\~{o}es (Brazil), and Korea Astronomy and Space Science Institute (Republic of Korea).}
\blfootnote{Magellan telescope time was granted by \acs{NSF}’s \acs{NOIRLab}, through the \ac{TSIP}. \ac{TSIP} was funded by \ac{NSF}.}

\section{Introduction}
\label{282P:introduction}

Volatiles are vital to life as we know it and are critically important to future space exploration, yet basic knowledge about where volatiles (e.g., H$_2$O, CO, CH$_4$) are located within our own solar system is still incomplete. Moreover, the origin of solar system volatiles, including terrestrial water, remains inconclusive. Investigating sublimation-driven active solar system bodies can help answer these questions \citep{hsiehPopulationCometsMain2006,jewittAsteroidCometContinuum2022}.

We define volatile reservoirs as a dynamical class of minor planet that harbors volatile species, such as water ice. Comets have long been known to contain volatiles, but other important reservoirs are coming to light, such as the active asteroids -- objects on orbits normally associated with asteroids, such as those found in the main-belt, that surprisingly display cometary features such as tails and/or comae \citep{jewittActiveAsteroids2015a}. Fewer than 30 active asteroids have been discovered \citep{chandlerSAFARISearchingAsteroids2018} since the first, (4015)~Wilson-Harrington, was discovered in 1949 \citep{cunninghamPeriodicCometWilsonHarrington1950} and, as a result, they remain poorly understood.

One scientifically important subset of active asteroids consists of members that display recurrent activity attributed to sublimation: the \acp{MBC} \citep{hsiehMainbeltCometsPanSTARRS12015}. An important diagnostic of indicator sublimating volatiles, like water ice, is recurrent activity near perihelion \citep{hsiehOpticalDynamicalCharacterization2012,snodgrassMainBeltComets2017}, a feature common to the \acp{MBC} \citep{hsiehMainbeltCometsPanSTARRS12015,agarwalBinaryMainbeltComet2017,hsieh2016ReactivationsMainbelt2018}. Fewer than 10 recurrently active \acp{MBC} have been discovered (though others exhibit activity attributed to sublimation), and as a result we know very little about this population.

Another potential volatile reservoir, active Centaurs, came to light after comet 29P/Schwassmann-Wachmann 1 \citep{schwassmannNEWCOMET1927} was identified as a Centaur following the 1977 discovery of (2060)~Chiron \citep{kowalSlowMovingObjectKowal1977}. Centaurs, found between the orbits of Jupiter and Neptune, are cold objects thought to primarily originate in the Kuiper Belt prior to migrating to their current orbits (see review, \citealt{jewittActiveCentaurs2009}). The dynamical properties of these objects are discussed in Section \ref{282P:sec:dynamicalClassification}. 
Fewer than 20 active Centaurs have been discovered to date, thus they, like the active asteroids, are both rare and poorly understood.


In order to enable the study of active objects in populations not typically associated with activity (e.g., \acp{NEO}, main-belt asteroids), we created a Citizen Science project designed to identify roughly 100 active objects via volunteer identification of activity in images of known minor planets. The Citizen Science paradigm involves concurrently crowdsourcing tasks yet too complex for computers to perform, while also carrying out an outreach program that engages the public in a scientific endeavor. Launched in Fall 2021, our \ac{NSF} funded, \acs{NASA} partner program \textit{Active Asteroids}\footnote{\url{http://activeasteroids.net}} immediately began yielding results.


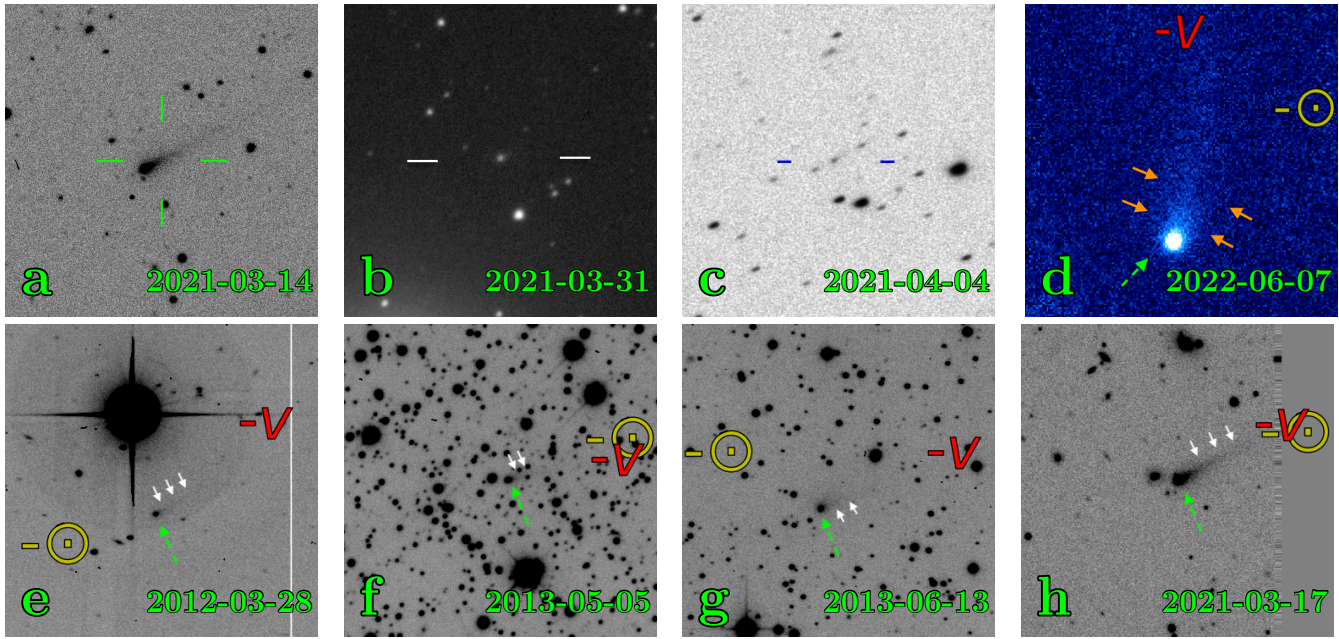
\begin{figure*}[ht]
    \centering
    \begin{tabular}{cccc}
    \begin{overpic}[width=0.23\linewidth]{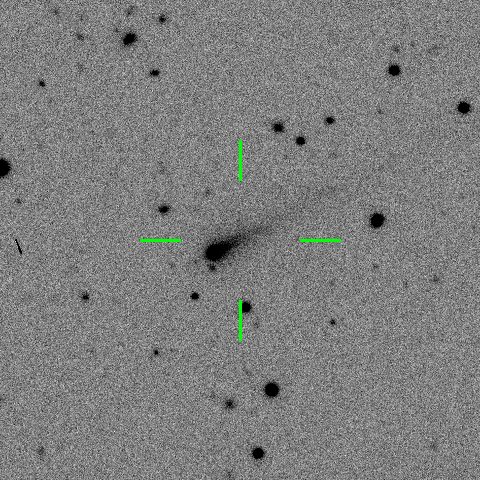}\put (5,7) {\huge\color{green} \textbf{\contour{black}{a}}}\put (45,8) {\large\color{green} \textbf{\contour{black}{2021-03-14}}}\end{overpic}  & 
    \begin{overpic}[width=0.23\linewidth]{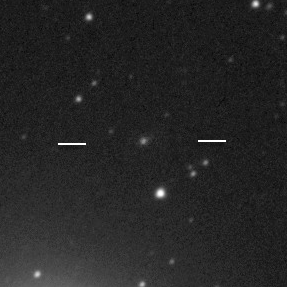}\put (5,7) {\huge\color{green} \textbf{\contour{black}{b}}}\put (45,8) {\large\color{green} \textbf{\contour{black}{2021-03-31}}}\end{overpic} &
    \begin{overpic}[width=0.23\linewidth]{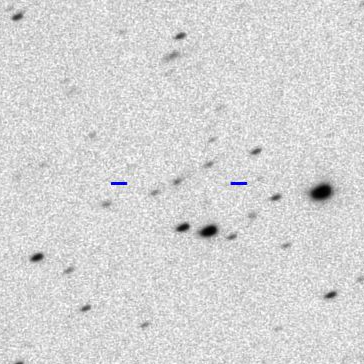}\put (5,7) {\huge\color{green} \textbf{\contour{black}{c}}}\put (45,8) {\large\color{green} \textbf{\contour{black}{2021-04-04}}}\end{overpic} &
    \begin{overpic}[width=0.23\linewidth]{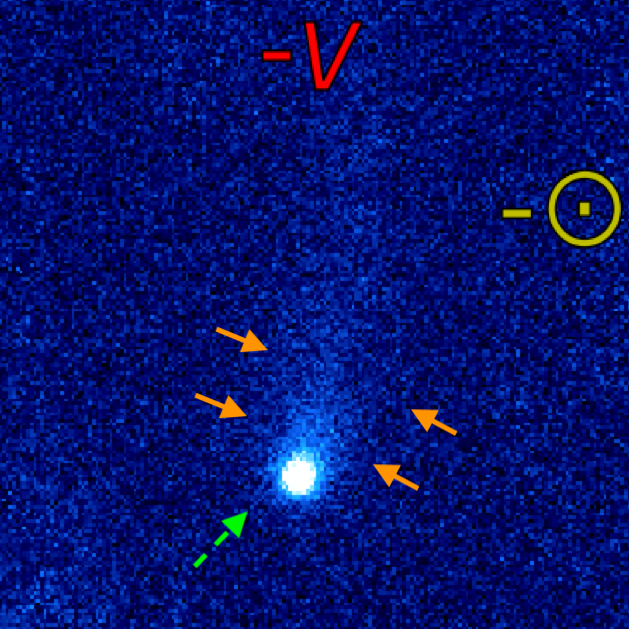}\put (5,7) {\huge\color{green} \textbf{\contour{black}{d}}}\put (45,8) {\large\color{green} \textbf{\contour{black}{2022-06-07}}}\end{overpic}\\
    \begin{overpic}[width=0.23\linewidth]{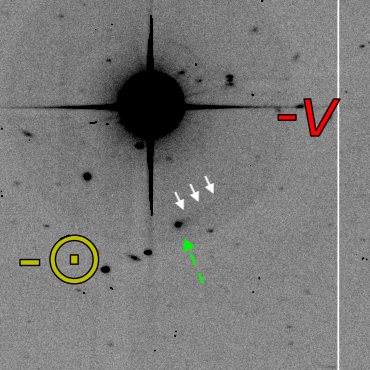}\put (5,7) {\huge\color{green} \textbf{\contour{black}{e}}}\put (45,8) {\large\color{green} \textbf{\contour{black}{2012-03-28}}}\end{overpic} & 
    \begin{overpic}[width=0.23\linewidth]{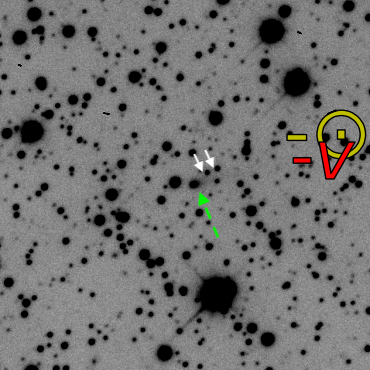}\put (5,7) {\huge\color{green} \textbf{\contour{black}{f}}}\put (45,8) {\large\color{green} \textbf{\contour{black}{2013-05-05}}}\end{overpic} &
    \begin{overpic}[width=0.23\linewidth]{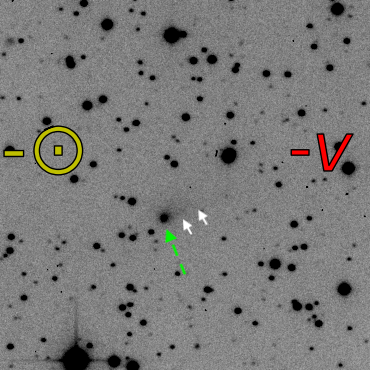}\put (5,7) {\huge\color{green} \textbf{\contour{black}{g}}}\put (45,8) {\large\color{green} \textbf{\contour{black}{2013-06-13}}}\end{overpic} &
    \begin{overpic}[width=0.23\linewidth]{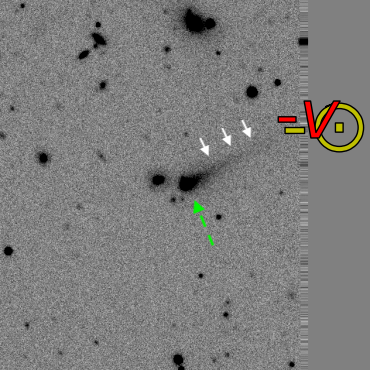}\put (5,7) {\huge\color{green} \textbf{\contour{black}{h}}}\put (45,8) {\large\color{green} \textbf{\contour{black}{2021-03-17}}}\end{overpic}
    \end{tabular}
    \caption{
    Top row: four images, spanning 15 months, showing \objnameFull{} activity during the recent 2021--2022 activity epoch. 
    \textbf{(a)} Epoch II thumbnail image of \objname{} was classified as ``active'' by 14 of 15 volunteers of our Citizen Science project \textit{Active Asteroids}, a NASA Partner program. This 90~s $i$ band image was taken with the Dark Energy Camera on UT 2021 March 14, Prop. ID 2019A-0305 (\acs{PI} Drlica-Wagner). 
    \textbf{(b)} Epoch II, 12$\times$300~s co-added exposures imaged by Michael Jäger with a QHY600 camera on a 14'' Newtonian telescope in Weißenkirchen, Austria. Image reproduced with permission of Michael Jäger. 
    \textbf{(c)} Epoch II 5$\times$300~s co-added images captured by Roland Fichtl using a CDS cooled Canon 5D Mark III camera on a 16'' Newtonian telescope in Engelhardsberg, Germany. Image reproduced with permission of Roland Fichtl. 
    \textbf{(d)} For this most recent Epoch II image we co-added six 120~s $g'$ band images of \objname{} (green dashed arrow) we acquired on UT 7 June 2022 with the \ac{GMOS} imager on the 8.1~m Gemini South telescope (Prop. ID GS-2022A-DD-103, \acs{PI} Chandler); a tail is clearly visible (orange arrows).
    Bottom row: Archival images of \objname{} that show clear evidence of activity. For each 126\arcsec$\times$126\arcsec thumbnail image, north is up and east is left. With the center of each image as the origin, the antisolar (yellow -$\odot$) and antimotion (red -$v$) directions (often correlated with tail appearance) are indicated. \objname{} is indicated by the green dashed arrow, and visible activity is marked by the white arrows.
    \textbf{(e)} Epoch I image from UT 2012 March 28 MegaPrime 120~s $r$ band, Prop. ID 12AH16 (\acs{PI} Wainscoat). 
    \textbf{(f)} Epoch I image from UT 2013 May 5 DECam 150~s $r$ band, Prop. ID 2013A-0327 (\acs{PI} Rest). 
    \textbf{(g)} Epoch I image from UT 2013 June 13 MegaPrime 120~s $r$ band, Prop. ID 13AH09 (\acs{PI} Wainscoat). 
    \textbf{(h)} Epoch II image from UT 2021 March 17 DECam 90~s $i$ band, Prop. ID 2019A-0305 (\acs{PI} Drlica-Wagner). 
    } 
    \label{282P:fig:282P}
\end{figure*}

\objnameFull{}, hereafter \objname{}, was originally discovered as 2003~BM$_{80}$ on UT 2003 Jan 31 by Brian Skiff of the \ac{LONEOS} survey, and independently as 2003~FV$_{112}$ by \ac{LINEAR} on UT 2003 Apr 18. \objname{} was identified to be active during its 2012--2013 epoch (centered on its perihelion passage) in 2013 \citep{bolinComet2003BM2013}. 
Here, we introduce an additional activity epoch, spanning 2021--2022.

In this work we (1) present our \ac{NASA} Partner Citizen Science project \textit{Active Asteroids}, (2) describe how volunteers identified activity that led to our investigation into \objname{}, (3) present (a) archival images and (b) new observations of \objname{} that show it has undergone periods of activity during at least two epochs (2012--2013 and 2021--2022) spanning consecutive perihelion passages, (4) classify \objname{} as a \ac{QHO}, (5) explore the migratory nature of this object through dynamical modeling, including identification of a dynamical pathway between \acp{QHO} and active asteroids, and (6) determine volatile sublimation as the most probable activity mechanism.

\section{Citizen Science}
\label{282P:subsec:citsci}

We prepared thumbnail images (e.g., Figure \ref{282P:fig:282P}a) for examination by volunteers of our NASA Partner Citizen Science project \textit{Active Asteroids}, hosted on the Zooniverse\footnote{\url{https://www.zooniverse.org}} online Citizen Science platform. First we extract thumbnail images from publicly available pre-calibrated \ac{DECam} archival images using a pipeline, \ac{HARVEST}, first described in \cite{chandlerSAFARISearchingAsteroids2018} and expanded upon in \cite{chandlerSixYearsSustained2019,chandlerCometaryActivityDiscovered2020a,chandlerRecurrentActivityActive2021}. Each 126\arcsec$\times$126\arcsec\ thumbnail image shows one known minor planet at the center of the frame. We optimize the Citizen Science process by automatically excluding thumbnail images based on specific criteria, for example when (a) the image depth is insufficient for detecting activity, (b) no source was detected in the thumbnail center, and (c) too many sources were in the thumbnail to allow for reliable target identification; see \cite{chandlerChasingTailsActive2022} for an in-depth description.

Our workflow is simple: we show volunteers an image of a known minor planet and ask whether or not they see evidence of activity (like a tail or coma) coming from the object at the center of the image, as marked by a reticle (Figure \ref{282P:fig:282P}a). Each thumbnail is examined by at least 15 volunteers to minimize volunteer bias. 
To help train volunteers and validate that the project is working as intended, we created a training set of thumbnail images that we positively identified as showing activity, consisting of comets and other active objects, such as active asteroids. Training images are injected at random, though the interval of injection decays over time so that experienced volunteers only see a training image 5\% of the time.

We take the ratio of ``positive for activity'' classifications to the total number of classifications the object received, as a score to estimate the likelihood of the object being active. Members of the science team visually examines all images with a likelihood score of $\ge$80\% and flag candidates that warrant archival image investigation and telescope follow-up (Section \ref{282P:sec:observations}). We also learn of activity candidates through Zooniverse forums where users interact with each other, moderators, and our science team. Volunteers can share images they find interesting which has, in turn, led us directly to discoveries.

As of this writing, over 6,600 volunteers have participated in \textit{Active Asteroids}. They have conducted over 2.8$\times10^6$ classifications, completing assessment of over 171,000 thumbnail images. One image of \objname{} from UT 2021 March 14 (Figure \ref{282P:fig:282P}a) received a score of 93\% after 14 of 15 volunteers classified the thumbnail as showing activity. A second image from UT 2021 March 17 (Figure \ref{282P:fig:282P}h) was classified as active by 15 of 15 volunteers, providing additional strong evidence of activity from 2021 March.

\section{Observations}
\label{282P:sec:observations}


\subsection{Archival Data}
\label{282P:susbec:archivalData}

\begin{figure*}[ht]
	\centering
	\includegraphics[width=0.8\linewidth]{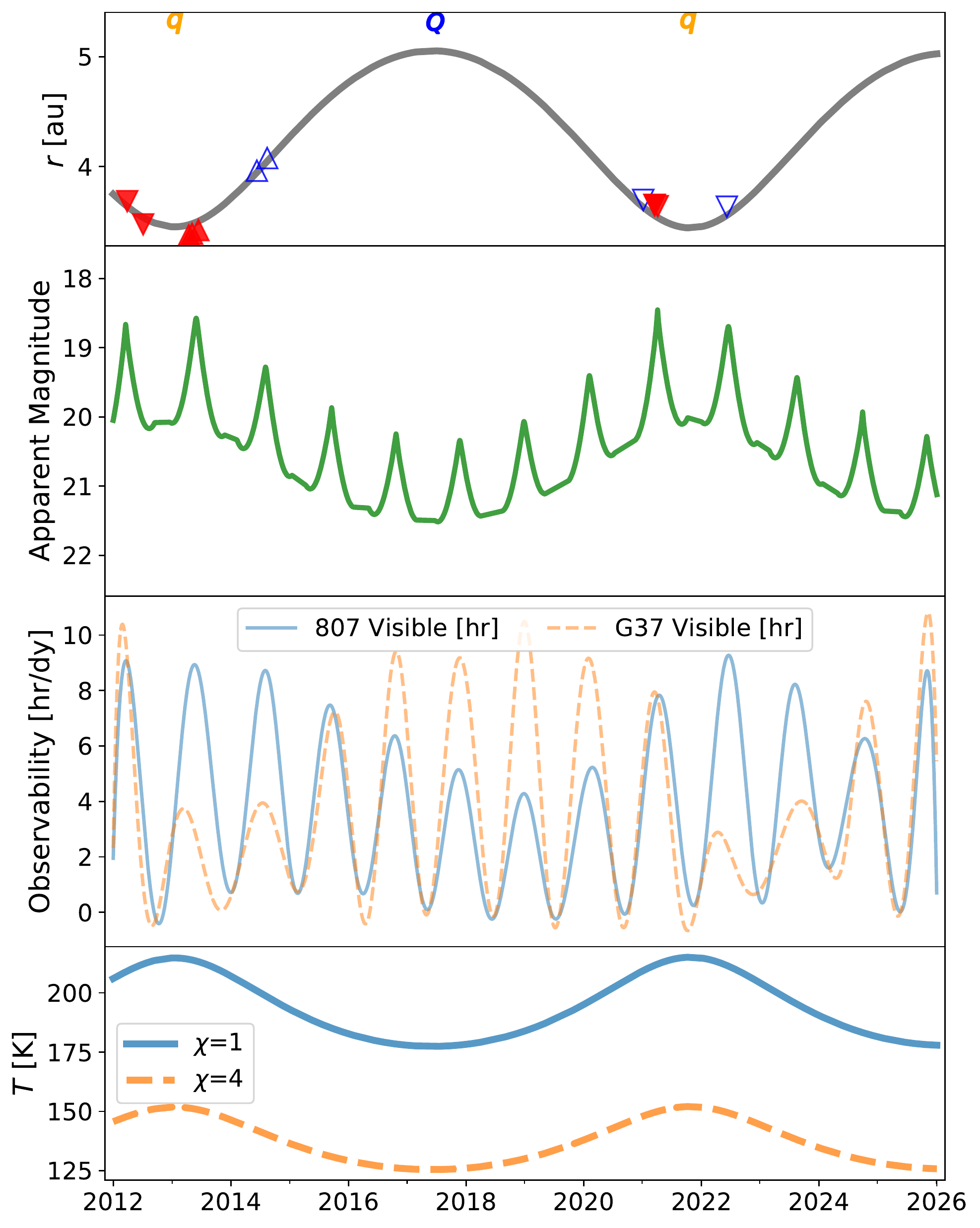}
	\caption{\objname{} heliocentric distance, observability, apparent brightness, observability, and temperature, from 2012 through 2025. 
	\textbf{Heliocentric Distance:} Activity detections (triangles) are marked as positive (filled red) and negative (unfilled blue) detections and as either
	inbound ($\blacktriangledown$) or outbound ($\blacktriangle$). 
	Observations are cataloged in Appendix \ref{282P:sec:observationsTable}. Also indicated are perihelion (orange $q$) and aphelion (blue $Q$) passages.
	\textbf{Apparent Magnitude:} The apparent $V$-band magnitude through time of \objname{}. 
	\textbf{Observability:} Our observability metric for \acf*{CTIO}, site code 807 (blue solid line) and the \acf*{LDT}, site code G37 (orange dashed line), depicting the number of hours \objname{} was observable ($>15\degr$ above the horizon between sunset and sunrise) during a given \ac{UT} observing date. Opposition events and conjunctions result in maxima and minima concurrent with apparent magnitude, respectively.
	\textbf{Temperature:} Modeled temperature by date for the thermophysical extremes: a ``flat slab'' ($\chi=1$, top line), and an isothermal body ($\chi=4$, bottom line).
	}
	\label{282P:fig:ActivityTimeline}
\end{figure*}

For each candidate active object stemming from \textit{Active Asteroids} we conduct an archival data investigation, following the procedure described in \cite{chandlerRecurrentActivityActive2021}. For this task, we query public astronomical image archives and identify images which may show \objname{} in the \ac{FOV}. We download the data, extract thumbnail images centered on \objname{}, and visually examine all images to search for evidence of activity.

After visually inspecting $>400$ thumbnail images we found 57 images (listed in Section \ref{282P:sec:observationsTable}) in which we could confidently identify \objname{} in the frame. The remaining images either did not probe faintly enough, did not actually capture \objname{} (e.g., \objname{} was not on a detector), or suffered from image artifacts that made the image unsuitable for activity detection. The 57 images span 22 observing dates; nine dates had at least one image we ascertained showed probable activity, five from the 2012--2013 epoch and four dates from the 2021--2022 apparition. 
Section \ref{282P:sec:observations} provides a complete listing of observations used in this work.

Figure \ref{282P:fig:ActivityTimeline} shows three plots with shared $x$-axes (years). 
%
Apparent magnitude and observability (the number of hours an object is above the horizon and the Sun is below the horizon) together provide insight into potential observational biases. For example, observations for detecting activity are ideal when \objname{} is brightest, near perihelion, and observable for many hours in an observing night. When contrasting hemispheres, this plot makes it clear that some periods (e.g., 2016 -- 2020) are more favorable for observations in the northern hemisphere, whereas other observation windows (e.g., 2013 -- 2015, 2022) are better suited to southern hemisphere facilities.

\subsection{Follow-up Telescope Observations}
\label{282P:subsec:telescopeobservations}

\paragraph{Magellan} During twilight on UT 2022 March 7 we observed \objname{} with the \ac{IMACS} instrument \citep{dresslerIMACSInamoriMagellanAreal2011} on the Magellan 6.5~m Baade telescope located atop Las Campanas Observatory (Chile). We successfully identified \objname{} in the images, however \objname{} was in front of a dense part of the Milky Way,
preventing us from unambiguously identifying activity. We used these observations to inform our Gemini \ac{SNR} calculations.

\paragraph{VATT} On UT 2022 April 6 we observed \objname{} with the 1.8~m \ac{VATT} at the \ac{MGIO} in Arizona (Proposal ID S165, \ac{PI} Chandler). \objname{} was in an especially dense part of the galaxy so we conducted test observations to assess the viability of activity detection under these conditions. We concluded object detection would be challenging and activity detection essentially impossible in such a dense field.

\paragraph{LDT} On UT 2022 May 21 we observed \objname{} with the \ac{LDT} in Arizona (PI: Chandler). Finding charts indicated \objname{} was in a less dense field compared to our \ac{VATT} observations, however we were hardly able to resolve \objname{} or identify any activity because the field was still too crowded.

\paragraph{Gemini South} On UT 2022 June 7 we observed \objname{} with the \ac{GMOS} South instrument \citep{hookGeminiNorthMultiObjectSpectrograph2004,gimenoOnskyCommissioningHamamatsu2016} on the 8.1~m Gemini South telescope located atop Cerro Pachón in Chile (Proposal ID GS-2022A-DD-103, \acs{PI} Chandler). We timed this observation to take place during a $\sim$10 day window when \objname{} was passing in front of a less dense region of the Milky Way. We acquired eighteen images, six each in $g'$, $r'$, and $i'$. Activity was clearly visible in the reduced data in all filters, with activity appearing strongest in $g'$ (Figure \ref{282P:fig:282P}d). Our observations confirmed \objname{} was still active, 15 months after the 2021 archival data, evidence supporting sublimation as the most likely cause for activity (Section \ref{282P:sec:mechanism}).

\section{Dynamical Modeling}
\label{282P:subsec:dynamicalmodeling}


We analyzed \objname{} orbital characteristics in order to (1) determine its dynamical class (Section \ref{282P:sec:dynamicalClassification}), and (2) inform our activity mechanism assessment (Section \ref{282P:sec:mechanism}). 
We simulated a cloud of 500 \objname{} orbital clones, randomly drawn from Gaussian distributions centered on the current fitted parameters of \objname{}, with widths corresponding to uncertainties of those fits (Appendix \ref{282P:sec:ObjectData} lists parameters and associated uncertainties), as reported by \acs{JPL} Horizons \citep{giorginiJPLOnLineSolar1996}.

We modeled the gravitational influence of the Sun and the planets (except Mercury) on each orbital clone using the \texttt{\ac{IAS15}} N-body integrator \citep{reinIAS15FastAdaptive2015}, typically accurate to machine precision, with the \texttt{REBOUND} \texttt{Python} package\footnote{\url{https://github.com/hannorein/rebound}} \citep{reinREBOUNDOpensourceMultipurpose2012,reinHybridSymplecticIntegrators2019}. We ran simulations 1,000 years forward and backward through time. Longer integrations were unnecessary because dynamical chaos ensues prior to $\sim$200 years ago and after $\sim$350 years into the future, thus no meaningful orbital elements can be derived outside of this window.


\begin{figure*}
    \centering
    \begin{tabular}{cc}
        \includegraphics[width=0.45\linewidth]{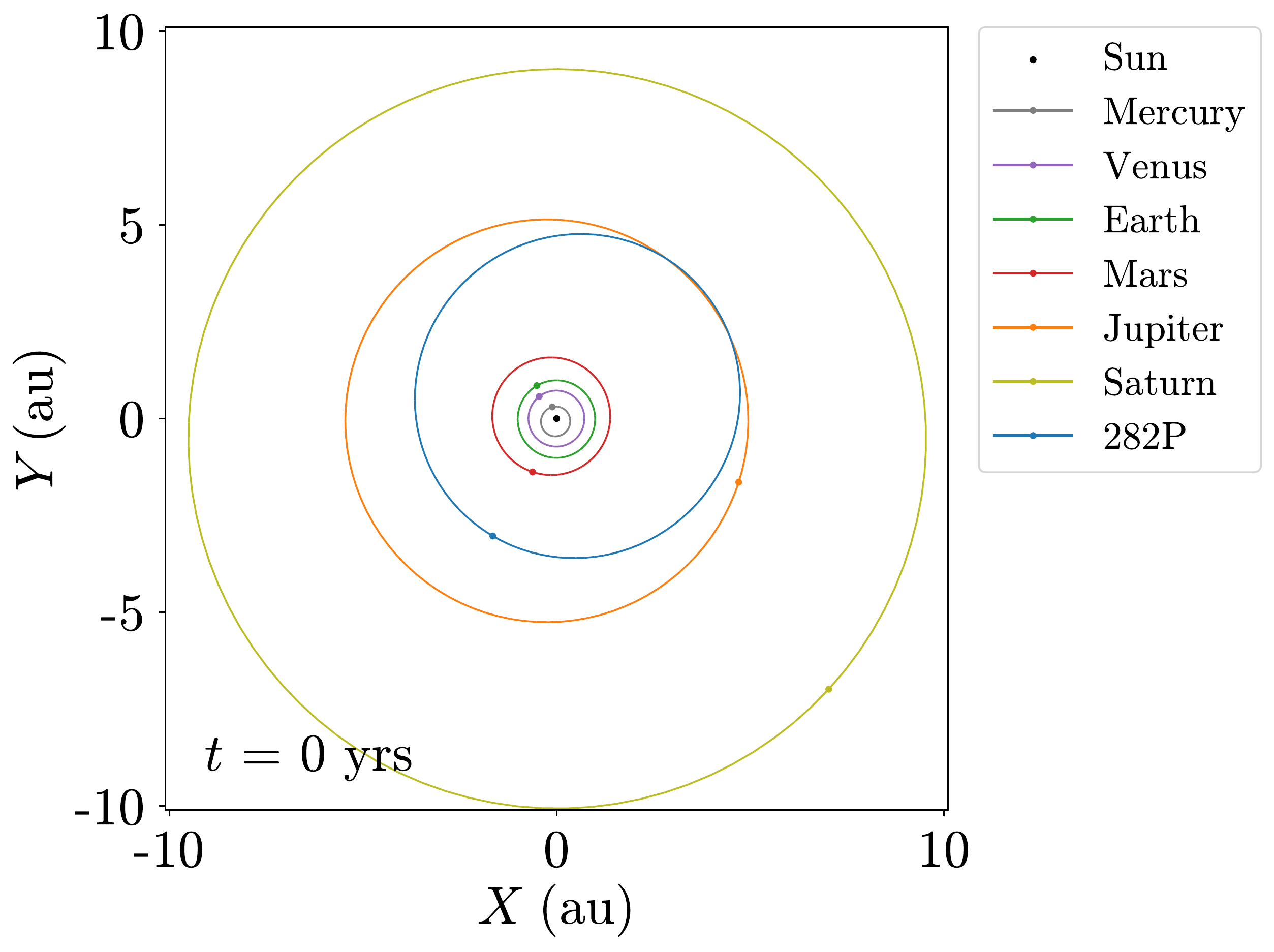} & \includegraphics[width=0.48\linewidth]{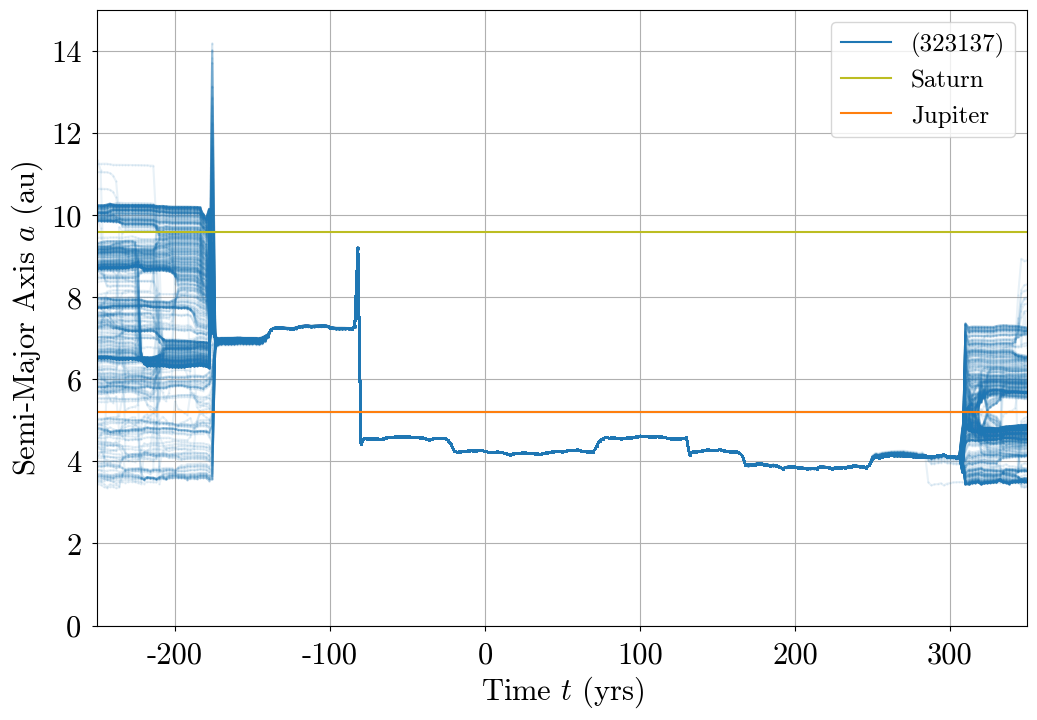}\\
        (a) & (b)\\
        \\
        \includegraphics[width=0.48\linewidth]{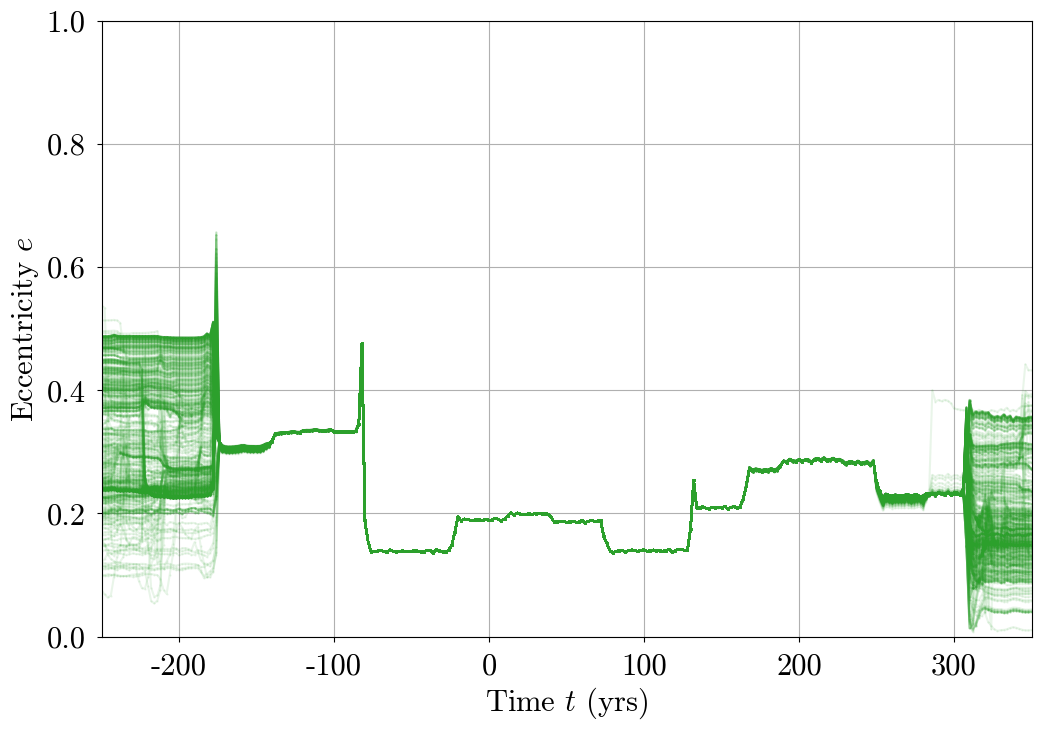} & \includegraphics[width=0.48\linewidth]{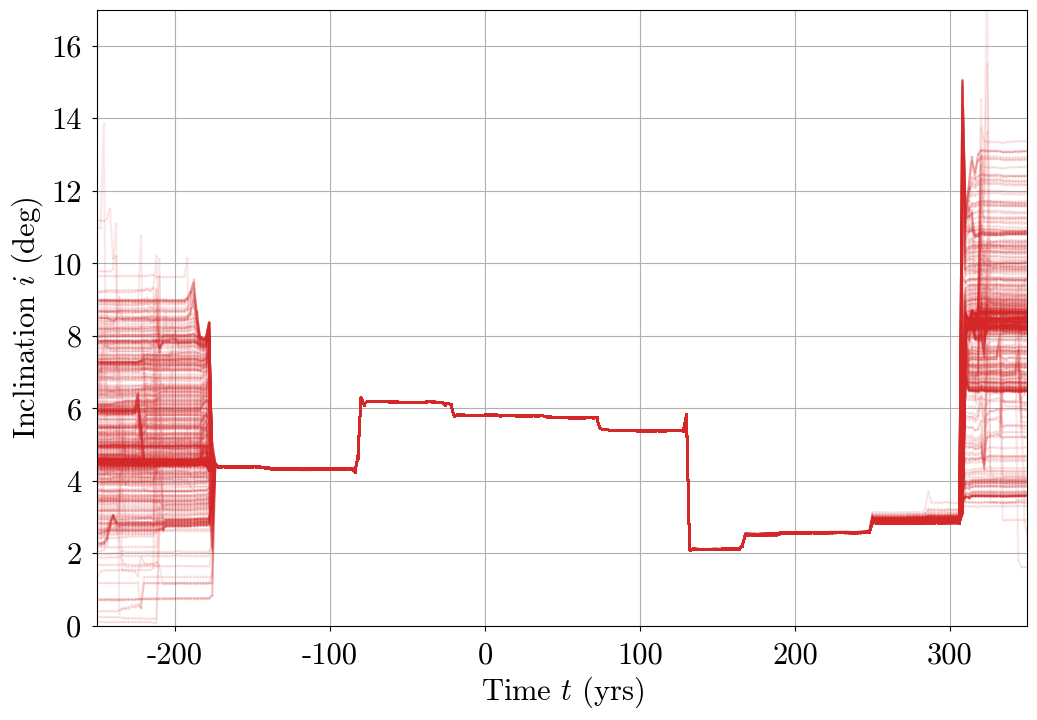}\\
       (c) & (d)\\
    \end{tabular}
    \caption{
        Results from dynamical integration of \objname{} orbital clones. For all plots, time $t=0$ corresponds to UT 2022 January 21. Jovian and Saturnian close encounters prevent accurate orbital parameter determination outside $-180\lesssim t\lesssim300$ yrs, given known orbital uncertainties.  
        \textbf{(a)} Orbital diagram for \objname{} and nearby planets; note that Uranus and Neptune were included in our simulations but they are not shown here. 
        \textbf{(b)} Semi-major axis $a$ evolution.
        \textbf{(c)} Eccentricity $e$ evolution. 
        \textbf{(d)} Inclination $i$ evolution. 
    }
    \label{282P:fig:orbitevolution1}
\end{figure*}

\begin{figure*}
    \centering
    \begin{tabular}{cc}
        \includegraphics[width=0.48\linewidth]{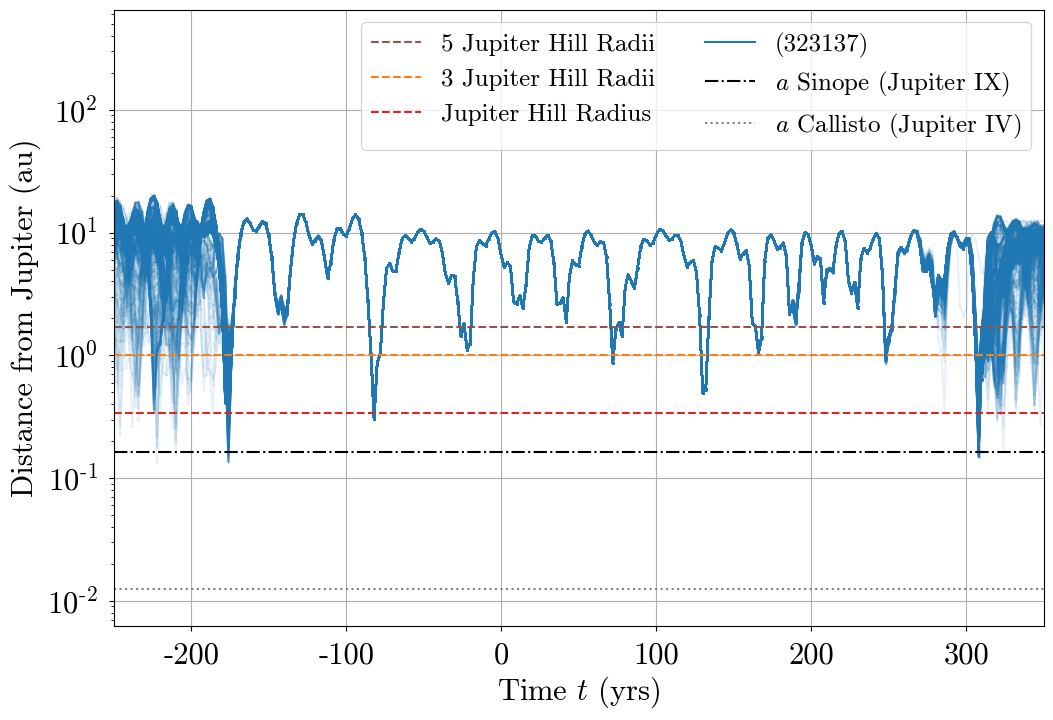} & \includegraphics[width=0.48\linewidth]{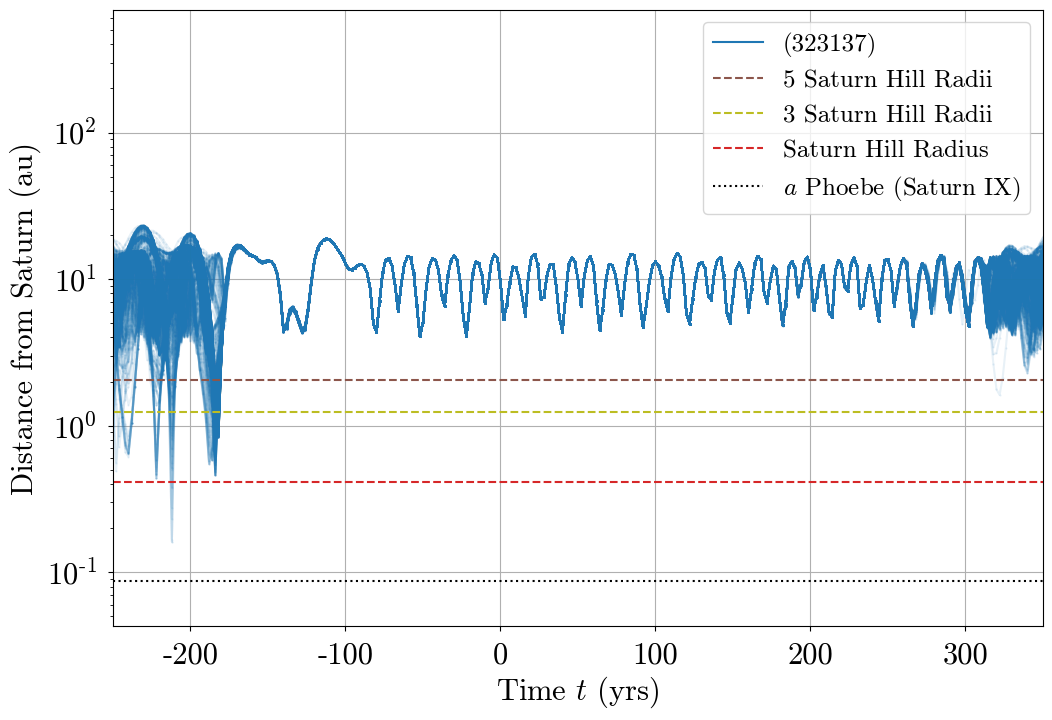}\\
        (a) & (b)\\
        \\
        \includegraphics[width=0.48\linewidth]{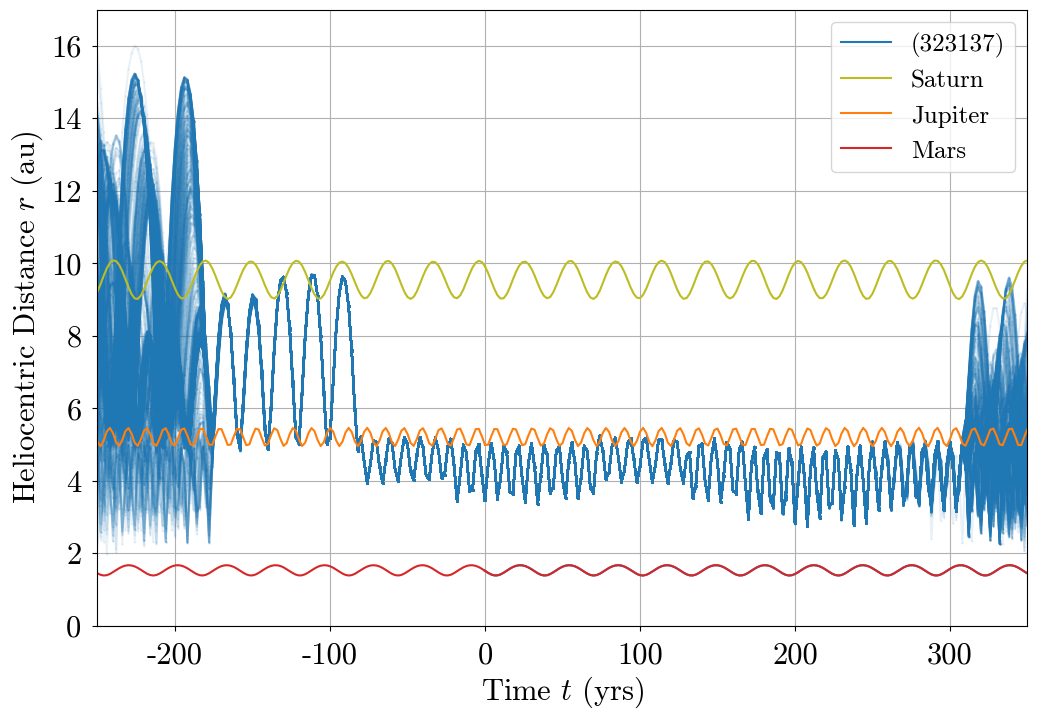} & \includegraphics[width=0.48\linewidth]{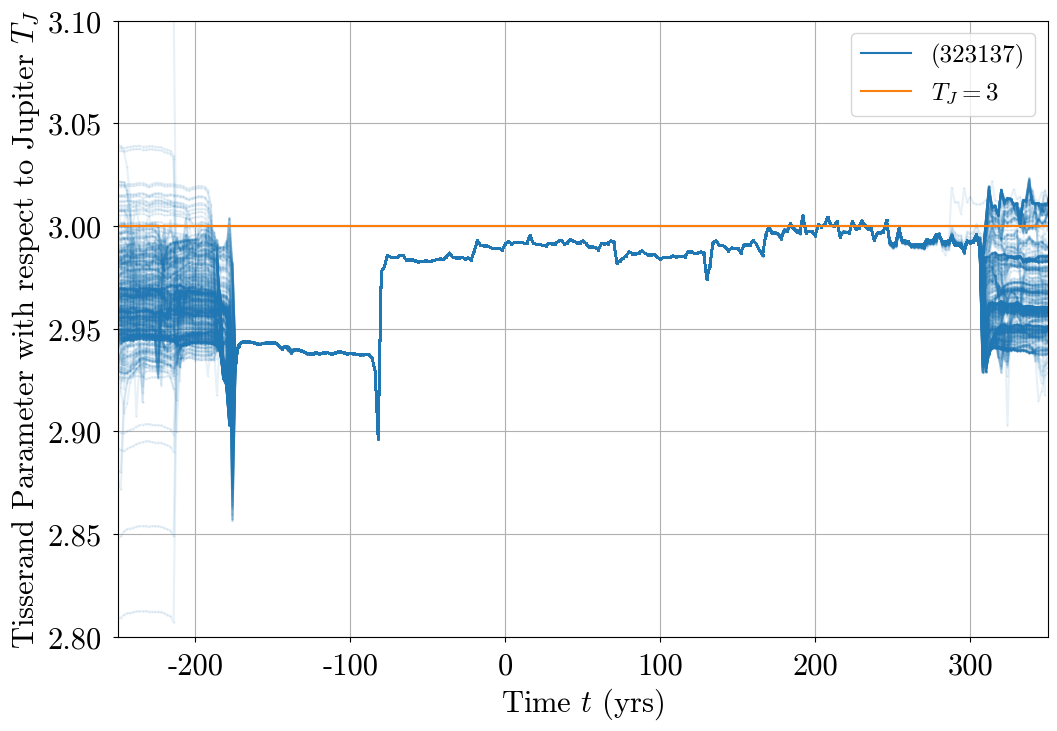}\\
        (c) & (d)\\
    \end{tabular}
    \caption{
    Additional results from dynamical integration of \objname{} orbital clones. For each plot, time $t=0$ is UT 2022 January 21. Close encounters with Jupiter and Saturn are so significant that orbital elements cannot be accurately determined before/after $-180\lesssim t\lesssim300$ yrs, given orbital uncertainties.
    \textbf{(a)} Distance between Jupiter and \objname{} as a function of time. Indicated Hill radii provide references for the degree of orbit alteration imparted by a close encounter. For reference, the semi-major axes of two Jovian moons are shown: Callisto, the outermost Galilean satellite, and Sinope \citep{nicholsonDiscoveryNinthSatellite1914}, a likely captured \citep{gravPhotometricSurveyIrregular2003a} distant irregular and retrograde Jovian moon.
    \textbf{(b)} Distance between Saturn and \objname{} as a function of time. The semi-major axis of the irregular Saturnian moon Phoebe, believed to be captured through close encounter \citep{johnsonSaturnMoonPhoebe2005,jewittIrregularSatellitesPlanets2007}, is given for reference.
    \textbf{(c)} Heliocentric distance $r$ evolution. 
    \textbf{(d)} Tisserand parameter with respect to Jupiter $T_\mathrm{J}$ (Equation \ref{282P:eq:TJ}), where the horizontal orange line representing $T_\mathrm{J}=3$ indicates the widely-adopted boundary between comet-like and asteroid-like orbits.
    }
    \label{282P:fig:orbitevolution2}
\end{figure*}

Results from the dynamical evolution of the \objname{} orbital clones are shown in Figure \ref{282P:fig:orbitevolution1} and Figure \ref{282P:fig:orbitevolution2}. For all plots, time $t=0$ corresponds to \ac{JD} 2459600.5 (UT 2022 Jan 21) and time ranges from $t=-250$ to $t=+350$ (1772--2372 AD). Horizontal lines at distances of one, three, and five Hill radii (Equation \ref{282P:eq:rH}) from Jupiter and Saturn are shown in Figure \ref{282P:fig:orbitevolution2} panels a and b. The Hill Radius \citep{hillResearchesLunarTheory1878} $r_H$ is a metric of orbital stability and indicates the region where a secondary body (e.g., a planet) has dominant gravitational influence over a tertiary body (e.g., a moon), with both related to a primary body, such as the Sun. At pericenter, the Hill radius of the secondary body can be approximated as

\begin{equation}
    r_\mathrm{H} \approx a(1-e)(m/3M)^{1/3},
    \label{282P:eq:rH}
\end{equation}

\noindent where $a$, $e$, and $m$ are the semi-major axis, eccentricity and mass of the secondary (Jupiter or Saturn in our case), respectively, and $M$ is the mass of the primary (here, the Sun). Close passages of a small body within a few Hill radii of a planet are generally considered to be significant perturbations and may drastically alter the orbit of the small body (see \citealt{hamiltonOrbitalStabilityZones1992} Section 2.1.2 for discussion). 

From $\sim$180 years ago until $\sim$300 years in the future, the orbit of \objname{} is well-constrained in our simulations. Figure \ref{282P:fig:orbitevolution2}a illustrates that \objname{} has roughly 10 close encounters (within $\sim$2 au) with Jupiter, and one with Saturn, over the range $-250<t<350$ yr. These encounters have a strong effect on the semi-major axis $a$ of \objname{} (Figure \ref{282P:fig:orbitevolution1}b), and, as illustrated by Figure \ref{282P:fig:orbitevolution2}d, a noticeable influence on its Tisserand parameter with respect to Jupiter $T_\mathrm{J}$, 

\begin{equation}
	T_\mathrm{J} = \frac{a_\mathrm{J}}{a} + 2\cos(i)\sqrt{\frac{a}{a_\mathrm{J}}\left(1-e^2\right)},
	\label{282P:eq:TJ}
\end{equation}

\noindent where $a_\mathrm{J}$ is the semi-major axis of Jupiter and $a$, $e$ and $i$ are the semi-major axis, eccentricity and inclination of the body, respectively. $T_\mathrm{J}$ essentially describes an object's close approach speed to Jupiter or, in effect, the degree of dynamical perturbation an object will experience as a consequence of Jovian influence. $T_\mathrm{J}$ is often described as invariant \citep{kresakJacobianIntegralClassificational1972} or conserved, meaning that changes in orbital parameters still result in the same $T_\mathrm{J}$, although, in practice, its value does change slightly as a result of close encounters (see Figure \ref{282P:fig:orbitevolution2}d).

Due to the small Jupiter-centric distances 
of \objname{} during these encounters, compounded by its orbital uncertainties, the past orbit of \objname{} (prior to $t\approx-180$ yrs) is chaotic. 
This dynamical chaos is plainly evident in all panels as orbital clones take a multitude of paths within the parameter space, resulting in a broad range of possible orbital outcomes due only to slight variations in initial \objname{} orbital parameters.

A consequential encounter with Saturn occurred around 1838 ($t\approx-184$~yr; Figure \ref{282P:fig:orbitevolution2}b), followed by another interaction with Jupiter in 1846 ($t=-176$ yr; Figure \ref{282P:fig:orbitevolution2}a). After these encounters \objname{} was a \ac{JFC} (100\% of orbital clones) with a semi-major axis between Jupiter's and Saturn's semi-major axes (Figure \ref{282P:fig:orbitevolution1}b), and crossing the orbits of both planets (Figure \ref{282P:fig:orbitevolution2}c). These highly perturbative passages placed \objname{} on the path that would lead to its current Quasi-Hilda orbit.

In 1940 ($t=-82$~yr), \objname{} had a very close encounter with Jupiter, at a distance of 0.3~au -- interior to one Hill radius. As seen in Figure \ref{282P:fig:orbitevolution1}a, this encounter dramatically altered \objname{}'s orbit, shifting \objname{} from an orbit primarily exterior to Jupiter to an orbit largely interior to Jupiter (Figure \ref{282P:fig:orbitevolution1}b). This same interaction also caused \objname{}'s orbit to migrate from Jupiter- and Saturn-crossing to only a Jupiter-crossing orbit (Figure \ref{282P:fig:orbitevolution2}c). This step in the orbital evolution of \objname{} also changed its $T_\mathrm{J}$ (Figure \ref{282P:fig:orbitevolution2}d) to be close to the traditional $T_\mathrm{J}=3$ comet--asteroid dynamical boundary. At this point in time, \objname{} remained a \ac{JFC} (100\% of orbital clones) despite its dramatic change in orbit. 

Around $t\approx200$ yr, \objname{} crosses the $T_\mathrm{J}=3$ boundary dividing the \ac{JFC}s and the asteroids on the order of 10 times. Although no major changes in the orbit \objname{} occur during this time, because of the stringency of this boundary, relatively minor perturbations result in oscillation between dynamical classes.

After a major encounter with Jupiter around 2330 AD ($t\approx308$ yrs), dynamical chaos again becomes dominant and remains so for the rest of the simulation. Following this encounter, the orbit of \objname{} does not converge around any single solution. Slight diffusion following the previous several Jupiter passages are also visible in Figure \ref{282P:fig:orbitevolution1}b-d and Figure \ref{282P:fig:orbitevolution2}a-d, and these also add uncertainty concerning encounters around 2301 to 2306 ($t\approx280$ to $285$ yrs). Although we are unable to precisely determine past and future orbits of \objname{} outside of $-180\lesssim t\lesssim300$ because of dynamical chaos, we are able to examine the fraction of orbital clones that finish the simulation (forwards and backwards) on orbits associated with different orbital classes.


\section{Dynamical Classifications: Past, Present and Future}
\label{282P:sec:dynamicalClassification}


Minor planets are often classified dynamically, based on orbital characteristics such as semi-major axis. 
%
\objname{} was labeled a \ac{JFC} by \cite{hsiehMainbeltCometsPanSTARRS12015}, in agreement with a widely adopted system that classifies objects dynamically based on their Tisserand parameter with respect to Jupiter, $T_\mathrm{J}$ (Equation \ref{282P:eq:TJ}).


Via Equation \ref{282P:eq:TJ}, Jupiter's $T_\mathrm{J}$ is 2.998 given $a_\mathrm{J}=5.20$, $e_\mathrm{J}=0.049$, and $i_\mathrm{J}=0.013$. Notably, objects with $T_\mathrm{J}>3$ cannot cross the Jovian orbit, thus their orbits are entirely interior or exterior to Jupiter's orbit \citep{levisonCometTaxonomy1996}. 
Objects with $T_\mathrm{J}<3$ are considered cometary \citep{levisonCometTaxonomy1996}, while those with $T_\mathrm{J}>3$ are not \citep{vaghiOriginJupiterFamily1973,vaghiOrbitalEvolutionComets1973}, a classification approach first suggested by \cite{carusiHighOrderLibrationsHalleyType1987,carusiCometTaxonomy1996}. \acp{JFC} have $2<T_\mathrm{J}<3$ (see e.g., \citealt{jewittActiveCentaurs2009}), 
and Damocloids and have $T_\mathrm{J}<2$ \citep{jewittFirstLookDamocloids2005}. 
We note, however, that the traditional $T_\mathrm{J}$ asteroid -- \ac{JFC} -- Damocloid continuum does not include (or exclude) \acp{QHO}.

As discussed in Section \ref{282P:introduction}, we adopt the \cite{jewittActiveCentaurs2009} definition of Centaur, which stipulates that a Centaur has an orbit entirely exterior to Jupiter, with both $q$ and $a$ interior to Neptune, and the body is not in 1:1 resonance with a planet. \objname{} has a semi-major axis $a=4.240$~au, well interior to Jupiter's $a_\mathrm{J}=5.2$~au. This disqualifies \objname{} as presently on a Centaurian orbit.

Active objects other than comets orbiting interior to Jupiter are primarily the active asteroids, defined as (1)  $T_\mathrm{J}>3$, (2) displaying comet-like activity, and (3) orbiting outside of mean-motion resonance with any of the planets. This last stipulation rules out the Jupiter Trojans (1:1 resonance) and the Hildas (3:2 resonance with Jupiter), even though both classes have members above and below the  $T_\mathrm{J}=3.0$ asteroid--comet transition line. We compute $T_\mathrm{J}=2.99136891\pm(3.73\times10^{-8})$ for \objname{} (see Appendix \ref{282P:sec:ObjectData} for a list of orbital parameters). 
These values do not exceed the traditional $T_\mathrm{J}=3$ cutoff; thus \objname{} cannot be considered an active asteroid in its current orbit. \acp{MBC} are an active asteroid subset defined as orbiting entirely within the main asteroid belt  \citep{hsiehMainbeltCometsPanSTARRS12015}. Figure \ref{282P:fig:orbitevolution2}c shows that \objname{}'s heliocentric distance does not stay within the boundaries of the Asteroid Belt (i.e., between the orbits of Mars and Jupiter), and so \objname{} does not qualify as a \ac{MBC}.

\begin{figure*}
    \centering
    \begin{tabular}{ccc}
         \includegraphics[width=0.32\linewidth]{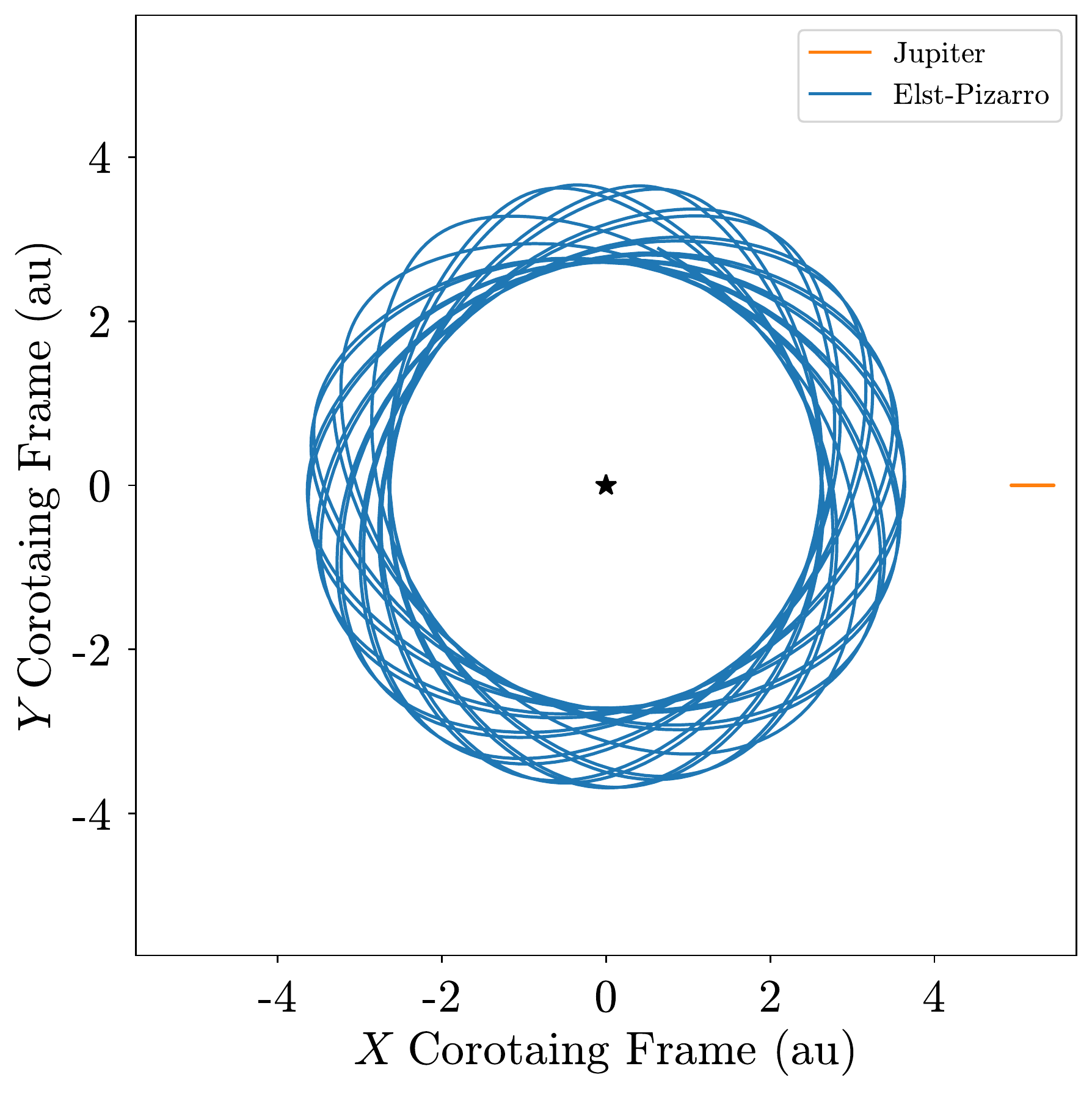} &
         \includegraphics[width=0.32\linewidth]{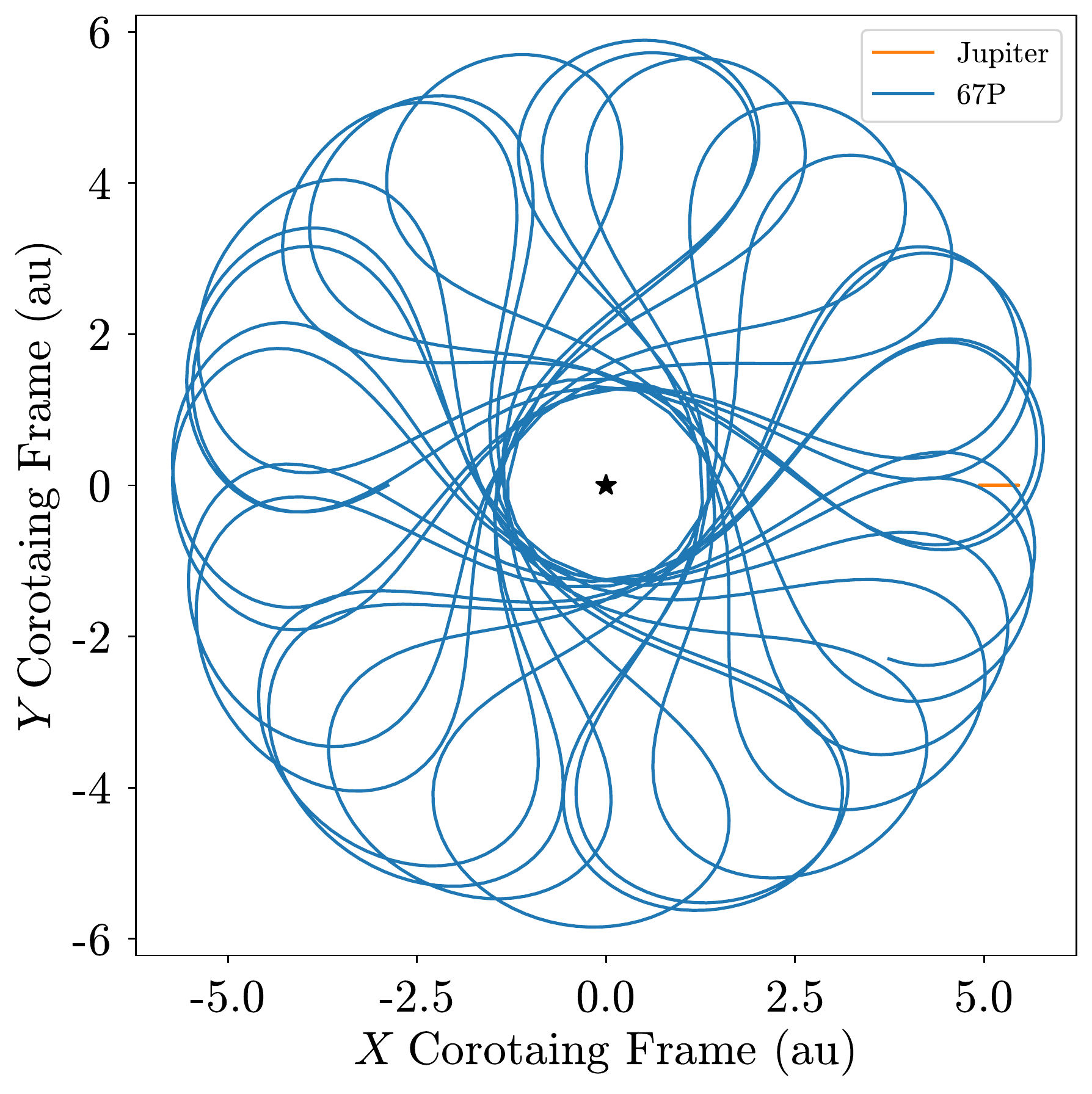} &
         \includegraphics[width=0.32\linewidth]{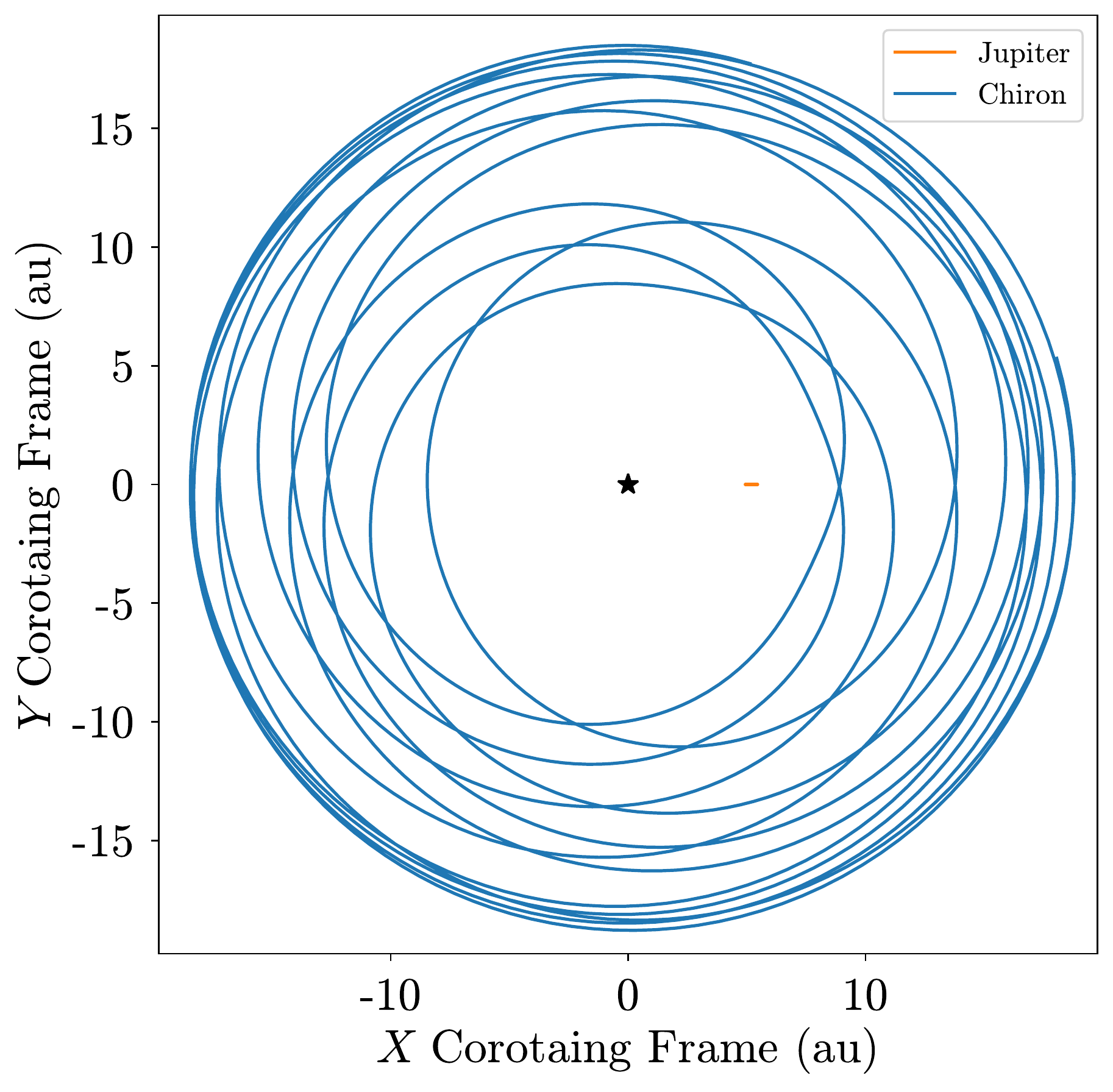}\\
         (a) Active Asteroid & (b) \acf{JFC} & (c) Centaur\\
          \\
          \includegraphics[width=0.32\linewidth]{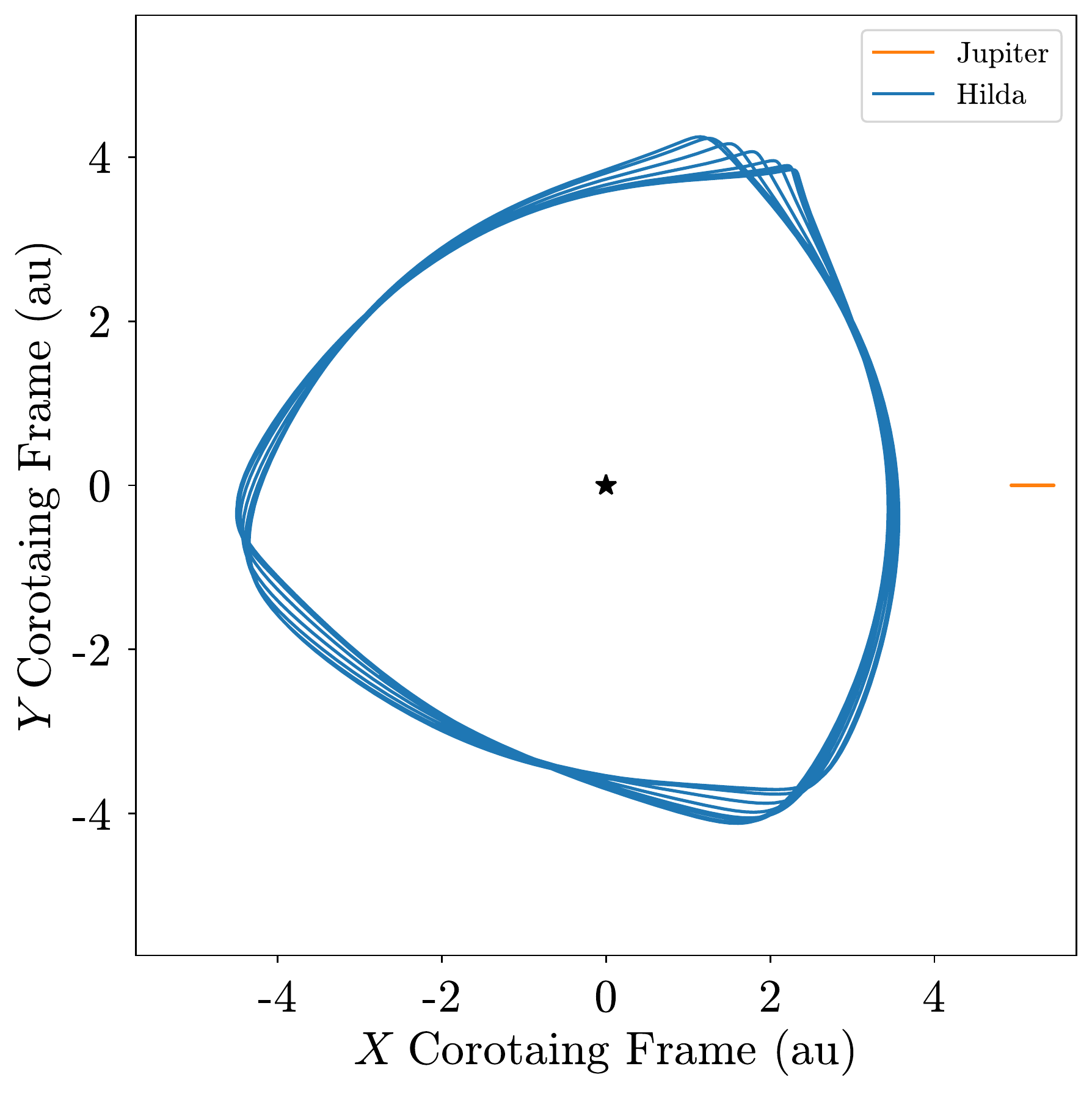} &
          \includegraphics[width=0.32\linewidth]{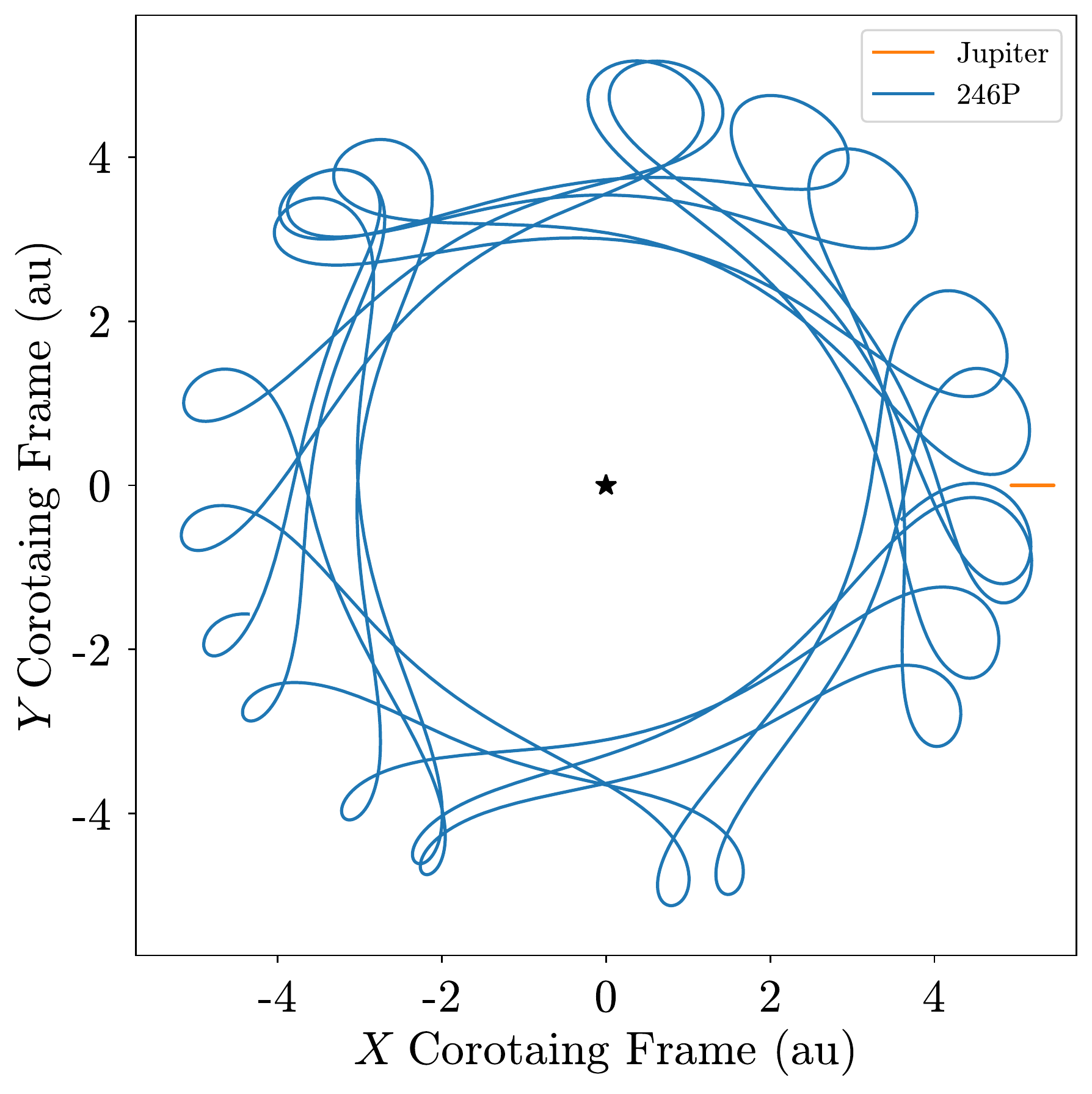} &
          \includegraphics[width=0.32\linewidth]{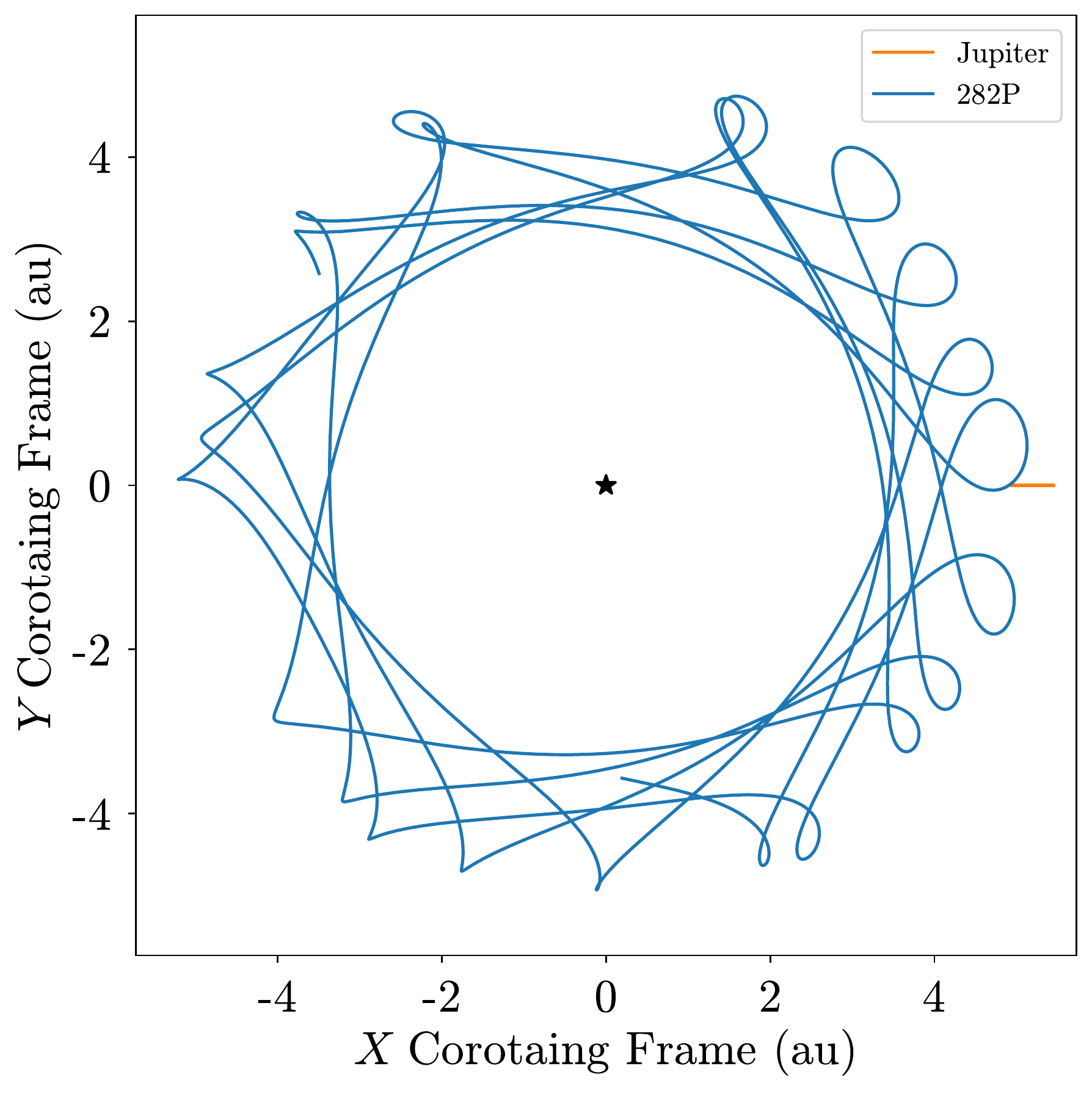} \\
          (d) Hilda & (e) Quasi-Hilda & (f) Quasi-Hilda\\
    \end{tabular}
    \caption{The orbital motion of minor planets (blue lines) as seen in the reference frame corotating with Jupiter (orange lines at right edge of plots). 
    (a) \ac{MBC} (7968)~Elst-Pizarro (133P).
    (b)  \ac{JFC} 67P/Churyumov-Gerasimenko (previously visited by the \ac{ESA} Rosetta Spacecraft).
    (c) Centaur (2060)~Chiron (95P).
    (d) (153)~Hilda, the namesake of the Hilda dynamical class, in the 3:2 interior mean-motion resonance with Jupiter. 
    (e) Quasi-Hilda 246P/\acs{NEAT}, also designated 2010~V$_2$ and 2004~F$_3$. 
    (f) Our object of study, \objname{}, in its Quasi-Hilda orbit.
    }
    \label{282P:fig:corotatingFrame}
\end{figure*}

Blurring the lines between \ac{JFC} and Hilda is the Quasi-Hilda regime. A Quasi-Hilda, also referred to as a \ac{QHO}, \ac{QHA} \citep{jewittOutburstingQuasiHildaAsteroid2020}, or \ac{QHC}, is a minor planet on an orbit similar to a Hilda \citep{tothQuasiHildaSubgroupEcliptic2006,gil-huttonCometCandidatesQuasiHilda2016}. Hildas are defined by their 3:2 interior mean-motion resonance with Jupiter, however Quasi-Hildas are not in this resonance, though they do orbit near it. Quasi-Hildas likely migrated from the \ac{JFC} region (see discussion, \citealt{jewittOutburstingQuasiHildaAsteroid2020}). We favor the term \ac{QHO} or \ac{QHA} over \ac{QHC}, given that fewer than 15 Quasi-Hildas have been found to be active, while the remainder of the $>270$ identified Quasi-Hildas \citep{gil-huttonCometCandidatesQuasiHilda2016} have not been confirmed to be active. Notable objects from the Quasi-Hilda class are 39P/Oterma \citep{otermaNEWCOMETOTERMA1942}, an object that was a Quasi-Hilda prior to 1963, when a very close (0.095~au) encounter with Jupiter redirected the object into a Centuarian orbit. Another notable Quasi-Hilda was D/Shoemaker-Levy~9, which famously broke apart and impacted Jupiter in 1994 (e.g., \citealt{weaverHubbleSpaceTelescope1995}).

Quasi-Hildas have orbital parameters similar to that of the Hildas, approximately $3.7 \lesssim a \lesssim 4.2$~au, $e\le0.3$, and $i\le20\degr$. In rough agreement, \objname{} has $a=4.24$~au, $e=0.188$, and $i=5.8\degr$ (Appendix \ref{282P:sec:ObjectData}). Hildas are also known for their trilobal orbits as viewed in the Jupiter corotating frame (caused by their residence in the 3:2 interior mean motion resonance with Jupiter), especially the namesake asteroid (153)~Hilda (Figure \ref{282P:fig:corotatingFrame}d). Because (153)~Hilda is in a stable 3:2 resonant orbit with Jupiter, its orbit remains roughly constant, with a small amount of libration over time. By contrast, Quasi-Hildas like 246P/\acs{NEAT} (Figure \ref{282P:fig:corotatingFrame}e) are near the same resonance and show signs of this characteristic trilobal pattern, however their orbits drift considerably on timescales of hundreds of years. \objname{} (Figure \ref{282P:fig:corotatingFrame}f) also displays a typical Quasi-Hilda orbit as viewed in the Jupiter corotating reference frame.

In the past, prior to 250~yr ago, 52\% (260) of the 500 orbital clones were \acp{JFC}, 48\% (239) were Cenaturs, 5\% (26) were already \acp{QHO}, and one (0.2\%) was an \ac{OMBA}. The most probable scenario prior to 250 years ago was that was either a \ac{JFC} or Centaur, both classes that trace their origins to the Kuiper Belt (see reviews, \citealt{morbidelliKuiperBeltFormation2020} and \citealt{jewittActiveCentaurs2009}, respectively).

In the future, after 350 years time, 81\% (403) of clones become \acp{JFC}, 18\% (90) remain \acp{QHO}, 14\% (69) become \acp{OMBA}, and 5.6\% (28) return to Centaurian orbits. Clearly the most likely scenario is that \objname{} will become a \ac{JFC}, however there are still significant possibilities that \objname{} remains a \ac{QHO} or becomes an active \ac{OMBA}.

\section{Thermodynamical Modeling}
\label{282P:sec:thermo}

In order to understand the approximate temperature ranges that \objname{} experiences over the course of its present orbit in order to (1) understand what role, if any, thermal fracture may play in the activity we observe, and (2) evaluate the likelihood of ices surviving on the surface, albeit with limited effect because of the narrow window ($\sim$500 years) of dynamically well-determined orbital parameters available (Section \ref{282P:subsec:dynamicalmodeling}).

Following the procedure of \cite{chandlerRecurrentActivityActive2021} (originally adapted from \citealt{hsiehMainbeltCometsPanSTARRS12015}), we compute the surface equilibrium temperature $T_\mathrm{eq}$ for \objname{} as a gray airless body. To accomplish this we begin with the water ice sublimation energy balance equation

\begin{equation}
{F_{\odot}\over r_h^2}(1-A) = \chi\left[{\varepsilon\sigma T_\mathrm{eq}^4 + L f_\mathrm{D}\dot m_{w}(T_\mathrm{eq})}\right]
\label{equation:sublim1}
\end{equation}

\noindent with the solar constant $F_{\odot}=1360$~W~m$^{-2}$,  heliocentric distance of the airless body $r_h$ (au), 
and the body's assumed Bond albedo is $A=0.05$; note that the true albedo could differ significantly from this value and thus it would be helpful to measure the albedo in the future when \objname{} is inactive. The heat distribution over the body is accounted for by $\chi$, which is bound by the coldest temperatures via the fast-rotating isothermal approximation ($\chi=1$), and the hottest temperatures via the ``slab'' sub-solar approximation, where one side of the object always faces the Sun. 
The assumed effective infrared emissivity is $\varepsilon=0.9$, $\sigma$ is the Stefan--Boltzmann constant, the latent heat of sublimation of water ice (which we approximate here as being independent of temperature) is $L=2.83$~MJ~kg$^{-1}$, 
the mantling-induced sublimation efficiency dampening is assumed to be $f_\mathrm{D}=1$ (absence of mantle), and the sublimation-driven water ice mass-loss rate in a vacuum $\dot m_\mathrm{w}$ is given by

\begin{equation}
\dot m_\mathrm{w} = P_\mathrm{v}(T) \sqrt{\mu\over2\pi k T}
\label{equation:sublim2}
\end{equation}

\noindent where the mass of one water molecule is $\mu=2.991\cdot 10^{-26}$~kg, 
$k$ is the Boltzmann constant, 

\noindent and the vapor pressure (in Pa) as a function of temperature $P_\mathrm{v}(T)$ is derived from the Clausius--Clapeyron relation,
\begin{equation}
P_\mathrm{v}(T) = 611 \times \exp\left[{{\Delta H_\mathrm{subl}\over R_g}\left({{1\over 273.16} - {1\over T}}\right)}\right]
\label{equation:sublim3}
\end{equation}

\noindent where the heat of sublimation for ice to vapor is $\Delta H_\mathrm{subl}=51.06$~MJ~kmol$^{-1}$, 
and the ideal gas constant is $R_g=8.314\times10^{-3}~\mathrm{MJ}~\mathrm{kmol}^{-1}$~K$^{-1}$.

Solving Equations \ref{equation:sublim1} -- \ref{equation:sublim3} for the body's heliocentric distance $r_\mathrm{h}$ (in au) as a function of equilibrium temperature $T_\mathrm{eq}$ and $\chi$ yields

\begin{equation} 
    r_\mathrm{h}(T_\mathrm{eq},\chi) = \frac{F_\odot\left(1-A\right)\chi^{-1}}{\epsilon\sigma T_\mathrm{eq}^4 + L f_\mathrm{D} \cdot 611\ e^{\frac{\Delta H_\mathrm{subl}}{R_\mathrm{G}}\left(\frac{1}{273.16\mathrm{K}} - \frac{1}{T_\mathrm{eq}}\right)}}
    \label{282P:eq:teq}
\end{equation}

We translate Equation \ref{282P:eq:teq} to a function of equilibrium temperature $T_\mathrm{eq}$ by computing $r_\mathrm{h}$ for an array of values (100~K to 300~K in this case), then fit a model to these data with a \texttt{SciPy} \citep{virtanenSciPyFundamentalAlgorithms2020} (\texttt{Python} package) univariate spline. Using this model we compute \objname{} temperatures for \objname{} heliocentric distances from perihelion and aphelion.

Figure \ref{282P:fig:ActivityTimeline} (bottom panel) shows the temperature evolution for the maximum and minimum solar heating distribution scenarios ($\chi=1$ and $\chi=4$, respectively) for \objname{} from 2012 through 2024. Temperatures range between roughly 175~K and 220~K for $\chi=1$, or 130~K and 160~K for $\chi=4$, with a $\sim45$~K maximum temperature variation in any one orbit. \objname{} spends some ($\chi=4$) or all ($\chi=1$) of its time with surface temperatures above 145~K. Water ice is not expected to survive above this temperature on Gyr timescales \citep{schorghoferLifetimeIceMain2008,snodgrassMainBeltComets2017}, however we showed in Section \ref{282P:subsec:dynamicalmodeling} that, prior to $\sim80$ years ago, \objname{} had a semi-major axis of $a>6$~au, a region much colder than 145~K. Even if \objname{} had spent most of its life with temperatures at the high end of our computed temperatures ($>220$~K), water ice can survive on Gyr timescales at shallow (a few cm) depths \citep{schorghoferLifetimeIceMain2008,prialnikCanIceSurvive2009}. Some bodies, such as (24)~Themis, have been found to have surface ices \citep{campinsWaterIceOrganics2010,rivkinDetectionIceOrganics2010} that suggest that an unknown mechanism may replenish surface ice with subsurface volatiles. In this case the ice lifetimes could be greatly extended.

\section{Activity Mechanism}
\label{282P:sec:mechanism}

Infrequent stochastic events, such as impacts (e.g., (596)~Scheila, \citealt{bodewitsCollisionalExcavationAsteroid2011,ishiguroObservationalEvidenceImpact2011,moreno596ScheilaOutburst2011}), are highly unlikely to be the activity mechanism given the multi-epoch nature of the activity we identified in this work. Moreover, it is unlikely that activity ceased during the 15 month interval between the UT 2021 March 14 archival activity and our UT 7 June 2022 Gemini South activity observations (Section \ref{282P:sec:observations}), when \objname{} was at a heliocentric distance $r_\mathrm{H}=3.548$~au and $r_\mathrm{H}$=3.556~au, respectively, and \objname{} was only closer to the Sun in the interim. Similarly, our archival data shows activity lasted $\sim15$ months during the 2012 -- 2013 apparition.

Recurrent activity is most commonly caused by volatile sublimation (e.g., 133P, \citealt{boehnhardtComet1996N21996,hsiehStrangeCase133P2004}) or rotational instability (e.g., (6478)~Gault, \citealt{kleynaSporadicActivity64782019,chandlerSixYearsSustained2019}). Rotational instability is impossible to rule out entirely for \objname{} because its rotation period is unknown. However, (1) no activity attributed to rotational stability for any object has been observed to be continuous for as long as the 15 month episodes we report, and (2) rotational instability is not correlated with perihelion passage. It is worth noting that there are not yet many known objects with activity attributed to rotational disruption, so it is still difficult to draw firm conclusions about the behavior of those objects. In any case it would be useful to measure a rotation period for \objname{} to help assess potential influence of rotational instability in the observed activity of \objname{}. The taxonomic class of \objname{} is unknown, but should \objname{} be classified as a member of a desiccated spectral class (e.g., S-type), then sublimation would not likely be the underlying activity mechanism. Color measurements or spectroscopy when \objname{} is quiescent would help determine its spectral class.

A caveat, however, is that many of our archival images were taken when \objname{} was significantly fainter than the images showing activity (Figure \ref{282P:fig:ActivityTimeline}), thereby making activity detection more difficult than if \objname{} was brighter. Consequently, archival images showing \objname{} were predominitely taken near its perihelion passage. The farthest evidently quiescent image of \objname{} was taken when it was at $\sim$4~au (Figure \ref{282P:fig:ActivityTimeline}). Thus we cannot state with total certainty that \objname{} was inactive elsewhere in its orbit.

Thermal fracture can cause repeated activity outbursts. For example, (3200)~Phaethon undergoes 600~K temperature swings, peaking at 800~K -- 1100~K, exceeding the serpentine-phyllosilicate decomposition threshold of 574~K \citep{ohtsukaSolarRadiationHeatingEffects2009}, and potentially causing thermal fracture \citep{licandroNatureCometasteroidTransition2007,kasugaObservations1999YC2008} including mass loss \citep{liRecurrentPerihelionActivity2013,huiResurrection3200Phaethon2017}. Temperatures on \objname{} reach at most $\sim220$~K (Figure \ref{282P:fig:ActivityTimeline}), with $\sim45$~K the maximum variation. Considering the relatively low temperatures and mild temperature changes we (1) consider it unlikely that \objname{} activity is due to thermal fracture, and (2) reaffirm that thermal fracture is generally considered a nonviable mechanism for any objects other than \acp{NEO}.

Overall, we find volatile sublimation on \objname{} the most likely activity mechanism, because (1) it is unlikely that an object originating from the Kuiper Belt such as \objname{} would be desiccated, (2) archival and new activity observations are from when \objname{} was near perihelion (Figure \ref{282P:fig:ActivityTimeline}), a characteristic diagnostic of sublimation-driven activity \citep[e.g.,][]{hsiehOpticalDynamicalCharacterization2012}, and (3) 15 months of continuous activity has not been reported for any other activity mechanism (e.g., rotational instability, impact events) to date, let alone two such epochs.

\section{Summary and Future Work}
\label{282P:sec:summary}

This study was prompted by Citizen Scientists from the NASA Partner program \textit{Active Asteroids} classifying two images of \objname{} from 2021 March as showing activity. Two additional images by astronomers Roland Fichtl and Michael Jäger brought the total number of images (from UT 2021 March 31 and UT 2021 April 4) to four. We conducted follow-up observations with the Gemini South 8.1~m telescope on UT 2022 June 7 and found \objname{} still active, indicating it has been active for $>15$ months during the current 2021 -- 2022 activity epoch. Our archival investigation revealed the only other known apparition, from 2012--2013, also spanned $\sim15$ months. Together, our new and archival data demonstrate \objname{} has been active during two consecutive perihelion passages, consistent with sublimation-driven activity.

We conducted extensive dynamical modeling and found \objname{} has experienced a series of $\sim5$ strong interactions with Jupiter and Saturn in the past, and that \objname{} will again have close encounters with Jupiter in the near future. These interactions are so strong that dynamical chaos dominates our simulations prior to 180 years ago and beyond 350 years in the future, but we are still able to statistically quantify a probable orbital class for \objname{} prior to $-180$ yr (52\% \acp{JFC}, 48\% Centaur) and after $+350$ yr (81\% \acp{JFC}, 18\% \ac{QHO}, 14\% \ac{OMBA}). We classify present-day \objname{} as a \acf{QHO}.

We carried out thermodynamical modeling that showed \objname{} undergoes temperatures ranging at most between 135~K and 220~K, too mild for thermal fracture but warm enough that surface water ice would not normally survive on timescales of the solar system lifetime. However, \objname{} arrived at its present orbit recently; prior to 1941 \objname{} was primarily exterior to Jupiter's orbit and, consequently, sufficiently cold for water ice to survive on its surface. Given that both activity apparitions (Epoch I: 2012 -- 2013 and Epoch II: 2021 -- 2022) each lasted over 15 months, and both outbursts spanned perihelia passage, we determine the activity mechanism to most likely be volatile sublimation.

Coma likely accounts for the majority of the reflected light we observe emanating from \objname{}, so it is infeasible to determine the color of the nucleus and, consequently, \objname{}'s spectral class (e.g., C-type, S-type). Measuring its rotational period would also help assess what (if any) role rotational instability plays in the observed activity. Specifically, a rotation period faster than the spin-barrier limit of two hours would indicate breakup.

Most images of \objname{} were taken when it was near perihelion passage (3.441~au), though there were observations from Epoch I that showed \objname{} clearly, without activity, when it was beyond $\sim$4~au. \objname{} is currently outbound and will again be beyond 4~au in mid-2023 and, thus, likely inactive; determining if/when \objname{} returns to a quiescent state would help bolster the case for sublimation-driven activity because activity occurring preferentially near perihelion, and a lack of activity elsewhere, is characteristic of sublimation-driven activity.

\objname{} is currently observable, especially from the southern hemisphere, however the object is passing in front of dense regions of the Milky Way until the end of 2022 November (see Lowell \texttt{AstFinder}\footnote{\url{https://asteroid.lowell.edu/astfinder/}} finding charts). \objname{} will be in a less dense region of the Milky Way and be observable, in a similar fashion to our Gemini South observations (Section \ref{282P:sec:observations}) on UT 2022 September 26 for $\sim$12 days, carefully timed for sky regions with fewer stars. As Earth's orbit progresses around the Sun, \objname{} becomes observable for less time each night through 2022 November, until UT 2022 December 26, when it becomes observable only during twilight. Observations during this window would help constrain the timeframe for periods of quiescence.

\section{Acknowledgements}
\label{282P:sec:acknowledgements}

The authors express their gratitude to the anonymous referee whose feedback improved the quality of this work a great deal.

We thank Dr.\ Mark Jesus Mendoza Magbanua of \ac{UCSF} for his frequent and timely feedback on the project. Many thanks for the helpful input from Henry Hsieh of the \ac{PSI} and David Jewitt of \ac{UCLA}.

We thank the \ac{NASA} Citizen Scientists involved in this work, with special thanks to moderator Elisabeth Baeten (Belgium) and our top classifier, Michele T. Mazzucato (Florence, Italy). Thanks also to super volunteers 
Milton K D Bosch MD (Napa, USA),
C. J. A. Dukes (Oxford, UK), 
Virgilio Gonano (Udine, Italy), 
Marvin W. Huddleston (Mesquite, USA), 
and 
Tiffany Shaw-Diaz (Dayton, USA), 
all of whom also classified images of \objname{}. Many thanks to additional classifiers of the three images of \objname{}: 
R. Banfield (Bad Tölz, Germany), 
@Boeuz (Penzberg, Germany), 
Dr. Elisabeth Chaghafi (Tübingen, Germany), 
Juli Fowler (Albuquerque, USA), 
M. M. Habram-Blanke (Heidelberg, Germany), 
@EEZuidema (Driezum, Netherlands), 
Brenna Hamilton (DePere, USA), 
Patricia MacMillan (Fredericksburg, USA), 
A. J. Raab (Seattle, USA), 
Angelina A. Reese (Sequim, USA), 
Arttu Sainio (Järvenpää, Finland), 
Timothy Scott (Baddeck, Canada), 
Ivan A. Terentev (Petrozavodsk, Russia), 
and 
Scott Virtes (Escondido, USA)
. 
Thanks also to \ac{NASA} Citizen Scientists 
Thorsten Eschweiler (Übach-Palenberg, Germany) 
and 
Carl Groat (Okeechobee, USA)
.

The authors express their gratitude to Prof. Mike Gowanlock (\acs{NAU}), Jay Kueny of \ac{UA} and Lowell Observatory, and the Trilling Research Group (\acs{NAU}), all of whom provided invaluable insights which substantially enhanced this work. Thank you William A. Burris (San Diego State University) for his insights into Citizen Science classifications. The unparalleled support provided by Monsoon cluster administrator Christopher Coffey (\acs{NAU}) and the High Performance Computing Support team facilitated the scientific process.

We thank Gemini Observatory Director Jennifer Lotz for granting our \ac{DDT} request for observations, German Gimeno for providing science support, and Pablo Prado for observing. Proposal ID GS-2022A-DD-103, \acs{PI} Chandler.

The VATT referenced herein refers to the Vatican Observatory’s Alice P. Lennon Telescope and Thomas J. Bannan Astrophysics Facility. We are grateful to the Vatican Observatory for the generous time allocations (Proposal ID S165, \acs{PI} Chandler). We especially thank Vatican Observatory Director Br. Guy Consolmagno, S.J. for his guidance, Vice Director for Tucson Vatican Observatory Research Group Rev.~Pavel Gabor, S.J. for his frequent assistance, Astronomer and Telescope Scientist Rev. Richard P. Boyle, S.J. for patiently training us to use the \ac{VATT} and for including us in minor planet discovery observations, Chris Johnson (\ac{VATT} Facilities Management and Maintenance) for many consultations that enabled us to resume observations, Michael Franz (\acs{VATT} Instrumentation) and Summer Franks (\ac{VATT} Software Engineer) for on-site troubleshooting assistance, and Gary Gray (\ac{VATT} Facilities Management and Maintenance) for everything from telescope balance to building water support, without whom we would have been lost.

This material is based upon work supported by the \acs{NSF} \ac{GRFP} under grant No.\ 2018258765. Any opinions, findings, and conclusions or recommendations expressed in this material are those of the author(s) and do not necessarily reflect the views of the \acl{NSF}. The authors acknowledge support from the \acs{NASA} Solar System Observations program (grant 80NSSC19K0869, PI Hsieh) and grant 80NSSC18K1006 (PI: Trujillo).

Computational analyses were run on Northern Arizona University's Monsoon computing cluster, funded by Arizona's \ac{TRIF}. This work was made possible in part through the State of Arizona Technology and Research Initiative Program. \acf{WCS} corrections facilitated by the \textit{Astrometry.net} software suite \citep{langAstrometryNetBlind2010}.

This research has made use of data and/or services provided by the \ac{IAU}'s \ac{MPC}. 
This research has made use of \acs{NASA}'s Astrophysics Data System. 
This research has made use of The \acf{IMCCE} SkyBoT Virtual Observatory tool \citep{berthierSkyBoTNewVO2006}. 
This work made use of the \texttt{FTOOLS} software package hosted by the \acs{NASA} Goddard Flight Center High Energy Astrophysics Science Archive Research Center. 
\ac{SAO} \ac{DS9}: This research has made use of \texttt{\acs{SAO}Image\acs{DS9}}, developed by \acl{SAO} \citep{joyeNewFeaturesSAOImage2006}. \acf{WCS} validation was facilitated with Vizier catalog queries \citep{ochsenbeinVizieRDatabaseAstronomical2000} of the Gaia \ac{DR} 2 \citep{gaiacollaborationGaiaDataRelease2018} and the \acf{SDSS DR-9} \citep{ahnNinthDataRelease2012} catalogs. 
This work made use of AstOrb, the Lowell Observatory Asteroid Orbit Database \textit{astorbDB} \citep{bowellPublicDomainAsteroid1994,moskovitzAstorbDatabaseLowell2021}. 
This work made use of the \texttt{astropy} software package \citep{robitailleAstropyCommunityPython2013}.

Based on observations at \ac{CTIO}, \acs{NSF}’s \acs{NOIRLab} (\acs{NOIRLab} Prop. ID 2019A-0305; \acs{PI}: A. Drlica-Wagner, \acs{NOIRLab} Prop. ID 2013A-0327; \acs{PI}: A. Rest), which is managed by the \acf{AURA} under a cooperative agreement with the \acl{NSF}. 
This project used data obtained with the \acf{DECam}, which was constructed by the \acf{DES} collaboration. Funding for the \acs{DES} Projects has been provided by the US Department of Energy, the US \acl{NSF}, the Ministry of Science and Education of Spain, the Science and Technology Facilities Council of the United Kingdom, the Higher Education Funding Council for England, the National Center for Supercomputing Applications at the University of Illinois at Urbana-Champaign, the Kavli Institute for Cosmological Physics at the University of Chicago, Center for Cosmology and Astro-Particle Physics at the Ohio State University, the Mitchell Institute for Fundamental Physics and Astronomy at Texas A\&M University, Financiadora de Estudos e Projetos, Fundação Carlos Chagas Filho de Amparo à Pesquisa do Estado do Rio de Janeiro, Conselho Nacional de Desenvolvimento Científico e Tecnológico and the Ministério da Ciência, Tecnologia e Inovação, the Deutsche Forschungsgemeinschaft and the Collaborating Institutions in the Dark Energy Survey. The Collaborating Institutions are Argonne National Laboratory, the University of California at Santa Cruz, the University of Cambridge, Centro de Investigaciones Enérgeticas, Medioambientales y Tecnológicas–Madrid, the University of Chicago, University College London, the \acs{DES}-Brazil Consortium, the University of Edinburgh, the Eidgenössische Technische Hochschule (ETH) Zürich, Fermi National Accelerator Laboratory, the University of Illinois at Urbana-Champaign, the Institut de Ciències de l’Espai (IEEC/CSIC), the Institut de Física d’Altes Energies, Lawrence Berkeley National Laboratory, the Ludwig-Maximilians Universität München and the associated Excellence Cluster Universe, the University of Michigan, \acs{NSF}’s \acs{NOIRLab}, the University of Nottingham, the Ohio State University, the OzDES Membership Consortium, the University of Pennsylvania, the University of Portsmouth, \ac{SLAC} National Accelerator Laboratory, Stanford University, the University of Sussex, and Texas A\&M University.

These results made use of the \acf{LDT} at Lowell Observatory. Lowell is a private, non-profit institution dedicated to astrophysical research and public appreciation of astronomy and operates the \acs{LDT} in partnership with Boston University, the University of Maryland, the University of Toledo, \acf{NAU} and Yale University. The \acf{LMI} was built by Lowell Observatory using funds provided by the \acf{NSF} (AST-1005313).

\ac{VST} OMEGACam \citep{arnaboldiVSTVLTSurvey1998,kuijkenOmegaCAM16k16k2002,kuijkenOmegaCAMESONewest2011} data were originally acquired as part of the \ac{KIDS} \citep{dejongFirstSecondData2015}.

The \acs{Pan-STARRS}1 Surveys (PS1) and the PS1 public science archive have been made possible through contributions by the Institute for Astronomy, the University of Hawaii, the \acs{Pan-STARRS} Project Office, the Max-Planck Society and its participating institutes, the Max Planck Institute for Astronomy, Heidelberg and the Max Planck Institute for Extraterrestrial Physics, Garching, The Johns Hopkins University, Durham University, the University of Edinburgh, the Queen's University Belfast, the Harvard-Smithsonian Center for Astrophysics, the \ac{LCOGT} Network Incorporated, the National Central University of Taiwan, the \acl{STScI}, the \acl{NASA} under Grant No. NNX08AR22G issued through the Planetary Science Division of the \acs{NASA} Science Mission Directorate, the \acf{NSF} Grant No. AST-1238877, the University of Maryland, \ac{ELTE}, the Los Alamos National Laboratory, and the Gordon and Betty Moore Foundation.

Based on observations obtained with MegaPrime/MegaCam, a joint project of \ac{CFHT} and \ac{CEA}/\ac{DAPNIA}, at the \ac{CFHT} which is operated by the \acf{NRC} of Canada, the Institut National des Science de l'Univers of the \acf{CNRS} of France, and the University of Hawaii. The observations at the \acf{CFHT} were performed with care and respect from the summit of Maunakea which is a significant cultural and historic site.

Magellan observations made use of the \ac{IMACS} instrument \citep{dresslerIMACSInamoriMagellanAreal2011}.

This research has made use of the \acs{NASA}/\ac{IPAC} \ac{IRSA}, which is funded by the \acl{NASA} and operated by the California Institute of Technology.

\vspace{5mm}
\facilities{
    Astro Data Archive, 
    Blanco (DECam),
    CFHT (MegaCam), 
    Gaia,
    Gemini-South (GMOS-S),
    IRSA, 
    LDT (LMI),
    Magellan: Baade (TSIP),
    PO:1.2m (PTF, ZTF), 
    PS1, 
    Sloan,
    VATT (VATT4K),
    VST (OmegaCAM)
}

\software{{\tt astropy} \citep{robitailleAstropyCommunityPython2013},
        {\tt astrometry.net} \citep{langAstrometryNetBlind2010},
        {\tt FTOOLS}\footnote{\url{https://heasarc.gsfc.nasa.gov/ftools/}},
        {\tt IAS15} integrator \citep{reinIAS15FastAdaptive2015},
        {\tt JPL Horizons} \citep{giorginiJPLOnLineSolar1996},
        {\tt Matplotlib} \citep{hunterMatplotlib2DGraphics2007},
        {\tt NumPy} \citep{harrisArrayProgrammingNumPy2020},
        {\tt pandas} \citep{mckinneyDataStructuresStatistical2010,rebackPandasdevPandasPandas2022},
        {\tt REBOUND} \citep{reinREBOUNDOpensourceMultipurpose2012,reinHybridSymplecticIntegrators2019},
        {\tt SAOImageDS9} \citep{joyeNewFeaturesSAOImage2006},
        {\tt SciPy} \citep{virtanenSciPyFundamentalAlgorithms2020},
        {\tt Siril}\footnote{\url{https://siril.org}},
        {\tt SkyBot} \citep{berthierSkyBoTNewVO2006},
        {\tt termcolor}\footnote{\url{https://pypi.org/project/termcolor}},
        {\tt tqdm} \citep{costa-luisTqdmFastExtensible2022},
        {\tt Vizier} \citep{ochsenbeinVizieRDatabaseAstronomical2000}
          }

\clearpage
\appendix

\section{Table of Observations}
\label{282P:sec:observationsTable}

\begin{center}
\footnotesize

\begin{tabular}{cccccrccccrrc}
Figure$^\mathrm{a}$ & Act.$^\mathrm{b}$ & Obs. Date$^\mathrm{c}$   & Source              & $N^\mathrm{d}$  & Exp. [s]$^\mathrm{e}$ & Filter(s) & V$^\mathrm{f}$    & $r$ [au]$^\mathrm{g}$    & STO [$\degr$]$^\mathrm{h}$  & $\nu$ [$\degr$]$^\mathrm{i}$    & \%$_{Q\rightarrow q}^\mathrm{j}$ & Note$^\mathrm{k}$\\
\hline\hline
                                &   & 04 Feb 2011 & PS1             & 2     &   40  & $r$       & 19.7 & 4.26 &  3.3 & 258.1 & 84\% & \ref{20120224}  \\
                                &   & 16 Feb 2012 & PS1             & 1     &   45  & $i$       & 19.3 & 3.69 &  8.9 & 306.4 & 95\% & \ref{20120224}  \\
                                &   & 24 Feb 2012 & PS1             & 1,1   &43, 40 & $g$, $r$  & 19.2 & 3.68 &  6.9 & 307.6 & 95\% & \obsnote{20120224}\ref{20120224}  \\
                                &   & 26 Feb 2012 & PS1             & 2     &   40  & $r$       & 19.2 & 3.67 &  6.3 & 307.9 & 96\% & \ref{20120224}\\
\ref{282P:fig:282P}e            & Y & 28 Mar 2012 & MegaPrime       & 2     &  120  & $r$       & 18.9 & 3.64 &  3.2 & 312.6 & 96\% & \obsnote{20120328}\ref{20120328}\\
                                & Y & 05 Jul 2012 & OmegaCAM        & 4     &  240  & $i$       & 20.1 & 3.54 & 16.1 & 328.0 & 98\% & \obsnote{20120705}\ref{20120705} \\
                                &   & 14 Apr 2013 & PS1             & 2     &   45  & $i$       & 19.3 & 3.47 & 12.8 &  14.9 & 99\% & \ref{20120224}\\ 
                                &   & 22 Apr 2013 & PS1             & 2     &   30  & $z$       & 19.2 & 3.47 & 11.1 &  16.3 & 99\% & \ref{20120224}\\
\ref{282P:fig:282P}f            & Y & 05 May 2013 & \acs{DECam}     & 2     &  150  & $r$       & 19.5 & 3.48 & 12.9 & 318.4 & 97\% & \obsnote{20130505}\ref{20130505}   \\
                                & Y & 15 May 2013 & PS1             & 1     &   43  & $g$       & 18.8 & 3.48 &  5.2 &  20.1 & 99\% & \ref{20120224}\\
\ref{282P:fig:282P}g            & Y & 13 Jun 2013 & MegaPrime       & 10    &  120  & $r$       & 18.8 & 3.50 &  4.9 &  24.8 & 99\% & \obsnote{20130613}\ref{20130613}  \\
                                &   & 03 Aug 2013 & PS1             & 2     &  80,60  & $y$, $z$  & 19.6 & 3.54 & 15.3 &  33.0 & 98\% & \ref{20120224}\\
                                &   & 11 Jun 2014 & PS1             & 2     &   45  & $i$       & 20.0 & 3.95 & 12.5 &  78.1 & 90\% & \ref{20120224}\\
                                &   & 14 Aug 2014 & PS1             & 3     &   45  & $i$       & 19.4 & 4.05 &  3.3 &  86.1 & 88\% & \ref{20120224}\\
                                &   & 15 Aug 2014 & PS1             & 4     &   45  & $i$       & 19.4 & 4.04 &  3.5 &  86.2 & 88\% & \ref{20120224}\\
                                &   & 04 Jan 2021 & \acs{ZTF}       & 1     &   30  & $r$       & 20.0 & 3.63 & 15.7 & 312.5 & 96\% & \obsnote{20210104}\ref{20210104}\\
                                &   & 07 Jan 2021 & \acs{ZTF}       & 1     &   30  & $g$       & 20.0 & 3.63 & 15.7 & 312.9 & 96\% & \ref{20210104}\\
                                &   & 09 Jan 2021 & \acs{ZTF}       & 1     &   30  & $r$       & 20.0 & 3.62 & 15.7 & 313.2 & 96\% & \ref{20210104}\\
\ref{282P:fig:282P}a            & Y & 14 Mar 2021 & \acs{DECam}     & 1     &   90  & $i$       & 18.9 & 3.55 &  6.1 & 323.1 & 98\% & \obsnote{20210314}\ref{20210314}    \\
\ref{282P:fig:282P}h            & Y & 17 Mar 2021 & \acs{DECam}     & 1     &   90  & $i$       & 18.9 & 3.55 &  5.2 & 323.6 & 98\% & \ref{20210314} \\
\ref{282P:fig:282P}b            & Y & 31 Mar 2021 & QHY600          & 1     & 2160  & UV/IR     & 18.5 & 3.54 &  0.7 & 325.9 & 98\% & \obsnote{20210331}\ref{20210331}\\
\ref{282P:fig:282P}c            & Y & 04 Apr 2021 & CDS-5D          & 1     & 1500  & (none)    & 18.5 & 3.54 &  0.5 & 326.4 & 98\% & \obsnote{20210404}\ref{20210404} \\
                                &   & 07 Mar 2022 & \acs{IMACS}     & 5     &   10  & WB4800-7800&20.0 & 3.48 & 16.3 &  22.3 & 99\% & \obsnote{20220307}\ref{20220307}\\
                                &   & 21 May 2022 & \acs{LDT}       & 3     &   90  & VR, $i$   & 19.1 & 3.54 &  8.3 &  34.4 & 98\% & \obsnote{20220521}\ref{20220521}\\
\ref{282P:fig:282P}d            & Y & 07 Jun 2022 & \ac{GMOS}-S     & 6,6,6 & 120   & $g$, $r$, $i$&18.8&3.56 &  3.8 &  37.2 & 98\% & \obsnote{20220607}\ref{20220607}\\
\end{tabular}
\end{center}
\noindent
$^\mathrm{a}$Figure showing the image. \\
$^\mathrm{b}$Activity identified in image(s). \\
$^\mathrm{c}$UT date of observation. \\
$^\mathrm{d}$Number of images. \\
$^\mathrm{e}$Exposure time. \\
$^\mathrm{f}$Apparent $V$-band magnitude (Horizons). \\
$^\mathrm{g}$Heliocentric distance. \\
$^\mathrm{h}$Sun--target--observer angle. \\
$^\mathrm{i}$True anomaly. \\
$^\mathrm{j}$Percentage to perihelion $q$ from aphelion $Q$, defined by $\%_{T\rightarrow q} = \left(\frac{Q - r}{Q-q}\right)\cdot 100\mathrm{\%}$. \\
$^\mathrm{k}$Note number. \\

\ref{20120224}: PS1 is the \acf{Pan-STARRS} One. 
\ref{20120328}: Prop. ID 12AH16, \acs{PI} Wainscoat. 
\ref{20120705}: Prop. ID 177.A-3016(D), \acs{PI} Kuijken. 
\ref{20130505}: \acf{DECam}; Prop. ID 2013A-0327, \acs{PI} Rest. 
\ref{20130613}: Prop. ID 13AH09, \acs{PI} Wainscoat. 
\ref{20210104}: \acf{ZTF}; Prop. ID 1467501130115, \acs{PI} Kulkarni; data acquired through \acs{ZTF} Alert Stream service \citep{pattersonZwickyTransientFacility2019}. 
\ref{20210314}: Prop. ID 2019A-0305, \acs{PI} Drlica-Wagner. 
\ref{20210331}: Michael Jäger (Weißenkirchen, Austria), QHY600 \ac{CCD} on a 14'' Newtonian, . 
\ref{20210404}: Roland Fichtl (Engelhardsberg, Germany), Central DS brand modified cooled Canon 5D Mark III on a 0.4~m f/2.5 Newtonian;  \url{http://www.dieholzhaeusler.de/Astro/comets/0282P.htm}. 
\ref{20220307}: \acf{IMACS}; PI Trujillo. 
\ref{20220521}: \acf{IMACS}; PI Trujillo. 
\ref{20220607}: \acf{GMOS}; Prop. ID GS-2022A-DD-103, \acs{PI} Chandler. 
\\

\clearpage
\section{Equipment and Archives}
\label{282P:sec:equipQuickRef}

\begin{center}

\footnotesize
    \begin{tabular}{llclccccc}
Instrument  & Telescope            & Pixel Scale    & Location                & \texttt{AstroArchive}         & \acs{ESO} & \acs{SSOIS}        & \acs{STScI}  & \acs{IRSA}\\
            &                       & [\arcsec/pix] &                   &   & & &\\
\hline
\hline
\acs{DECam}       & 4.0~m Blanco           & 0.263          & Cerro Tololo, Chile     & S,R &     &               S            &        \\
\acs{GMOS}-S        & 8.1~m Gemini South & 0.080        & Cerro Pachón, Chile &  &\\
\acs{IMACS}         & 6.5~m Baade       & 0.110          & Las Campanas, Chile & & &  \\ 
OmegaCAM    & 2.6~m \acs{VLT} Survey     & 0.214          & Cerro Paranal, Chile    &              & R   &           S            &        \\
GigaPixel1  & 1.8 m \acs{Pan-STARRS}1    & 0.258          & Haleakalā, Hawaii       &              &     &              S            & R      \\
\acs{LMI} & 4.3~m \acs{LDT}             & 0.120         & Happy Jack, Arizona       & & & \\
MegaPrime   & 3.6~m \acs{CFHT}           & 0.185          & Mauna Kea, Hawaii       &              &     &             S,R &        \\
\acs{PTF}/\acs{CFHT}12K & 48" Samuel Oschin    & 1.010          & Mt. Palomar, California &              &     &  &       &                 S,R   \\
\acs{ZTF} Camera  & 48" Samuel Oschin    & 1.012          & Mt. Palomar, California &              &     &  &       &                 S,R  \\
\acs{VATT}4K \acs{CCD}        & 1.8~m \acs{VATT}            & 0.188         & Mt. Graham, Arizona       & & & &\\
\end{tabular}
\raggedright
\footnotesize{\\
R indicates repository for data retrieval. S indicates search capability.\\
\texttt{AstroArchive}: \ac{NSF} \ac{NOIRLab} \texttt{AstroArchive} (\url{https://astroarchive.noirlab.edu}).\\
\ac{ESO}: \acl{ESO} (\url{https://archive.eso.org}).\\
\ac{IRSA}: \acs{NASA}/CalTech \ac{IRSA} (\url{https://irsa.ipac.caltech.edu}).\\
\acs{PTF}: The \ac{PTF}. 
\acs{SSOIS}: The \ac{SSOIS} (\citealt{gwynSSOSMovingObjectImage2012}, \url{https://www.cadc-ccda.hia-iha.nrc-cnrc.gc.ca/en/ssois/}).\\
\ac{STScI}: \url{https://www.stsci.edu/}.
}
\end{center}

This 
Table 
lists the instruments and telescopes used in this work, along with their respective pixel scales, locations, and data archives.

\clearpage
\section{282P/(323137) 2003 BM80 Data} 
\label{282P:sec:ObjectData}

We provide information current as of 2022 March 18 regarding \objnameFull{}{} below.

\begin{center}
    \small
	\begin{tabular}{lll}
		Parameter & Value & Source\\
		\hline\hline
		Designations & (323137), 2003~BM$_{80}$, 2003~FV$_{112}$, 282P & \acs{JPL} \acs{SBDB}, \acs{MPC}\\
		Discovery Date & 2003 January 31 & \acs{JPL} \acs{SBDB}, \acs{MPC}\\
		Discovery Observer(s) & \ac{LONEOS} & \acs{JPL} \acs{SBDB}, \acs{MPC}\\
		Discovery Observatory & Lowell Observatory & \acs{JPL} \acs{SBDB}, \acs{MPC}\\
		Discovery Site & Anderson Mesa Station, Arizona & \acs{JPL} \acs{SBDB}, \acs{MPC}\\
		Discovery Site Code & 688 & \acs{MPC} \\
		Activity Discovery Date & 2013 June 12 & \acs{CBET} 3559 \citep{bolinComet2003BM2013}\\
		Activity Discoverer(s) & Bryce Bolin, Larry Denneau, Peter Veres & \acs{CBET} 3559 \citep{bolinComet2003BM2013}\\
        Orbit Type & \acf{QHO} & this work\\ 
		Diameter & $D=$3.4$\pm$0.4~km & {\citet{harrisAsteroidsThermalInfrared2002}}\\ 
		Absolute $V$-band Magnitude & $H=13.63$ & \acs{MPC} (MPO648742)\\
        Geometric Albedo & Unknown & \\
        Assumed Geometric Albedo & 4\% & \cite{snodgrassSizeDistributionJupiter2011}\\
        Rotation Period & Unknown & \\
        Orbital Period & $P=8.732\pm(2.174\times10^{-7})$~yr & \acs{JPL} \acs{SBDB} \\ 
		Semi-major Axis & $a=4.240\pm(7.039\times10^{-8})$~au & \acs{JPL} \acs{SBDB}\\ 
		Perihelion Distance & $q=3.441\pm(3.468\times10^{-7})$~au & \acs{JPL} \acs{SBDB}\\ 
		Aphelion Distance & $Q=5.039\pm(8.366\times10^{-8})$~au & \acs{JPL} \acs{SBDB}\\ 
		Eccentricity & $e=0.188\pm(7.790\times10^{-8})$ & \acs{JPL} \acs{SBDB}\\ 
		Inclination & $i=5.812\degr\pm(1.166\degr\times10^{-5})$ & \acs{JPL} \acs{SBDB}\\ 
		Argument of Perihelion & $\omega=217.626\degr\pm(7.816\degr\times10^{-5})$ & \acs{JPL} \acs{SBDB}\\ 
		Longitude of Ascending Node & $\Omega=9.297\degr\pm(5.974\degr\times10^{-5})$ & \acs{JPL} \acs{SBDB}\\ 
		Mean Anomaly & $M=9.979\degr\pm(3.815\degr\times10^{-5})$ & \acs{JPL} \acs{SBDB}\\ 
		Tisserand Parameter w.r.t. Jupiter & $T_\mathrm{J}=2.99136891\pm\left(3.73\times10^{-8}\right)$ & this work\\
		Orbital Solution Date & 2021 October 8 & \acs{JPL} \acs{SBDB}\\
	\end{tabular}
	\raggedright\footnotesize

    Notes: 
    \ac{CBET} \footnote{\url{http://www.cbat.eps.harvard.edu}}. 
    \ac{JPL} \ac{SBDB} is the \acs{NASA} \acs{JPL} \acl{SBDB}\footnote{\url{https://ssd.jpl.nasa.gov/tools/sbdb_lookup.html}}. 
    \acs{MPC} is the \acl{MPC}\footnote{\url{https://minorplanetcenter.net}}. 
\end{center}



\clearpage
\section*{Acronyms}
\label{sec:acronyms}
\begin{acronym}
\acro{API}{Application Programming Interface}
\acro{APT}{Aperture Photometry Tool}
\acro{ARO}{Atmospheric Research Observatory}
\acro{AstOrb}{Asteroid Orbital Elements Database}
\acro{ASU}{Arizona Statue University}
\acro{AURA}{Association of Universities for Research in Astronomy}
\acro{BLT}{Barry Lutz Telescope}
\acro{CADC}{Canadian Astronomy Data Centre}
\acro{CASU}{Cambridge Astronomy Survey Unit}
\acro{CATCH}{Comet Asteroid Telescopic Catalog Hub}
\acro{CBAT}{Central Bureau for Astronomical Telegrams}
\acro{CCD}{charge-coupled device}
\acro{CEA}{Commissariat a l'Energes Atomique}
\acro{CBET}{Central Bureau for Electronic Telegrams}
\acro{CFHT}{Canada-France-Hawaii Telescope}
\acro{CNRS}{Centre National de la Recherche Scientifique}
\acro{CSBN}{Committee for Small Bodies Nomenclature}
\acro{CTIO}{Cerro Tololo Inter-American Observatory}
\acro{DART}{Double Asteroid Redirection Test}
\acro{DAPNIA}{Département d'Astrophysique, de physique des Particules, de physique Nucléaire et de l'Instrumentation Associée}
\acro{DECam}{Dark Energy Camera}
\acro{DES}{Dark Energy Survey}
\acro{DCT}{Discovery Channel Telescope}
\acro{DDT}{Director's Discretionary Time}
\acro{DR}{Data Release}
\acro{DS9}{Deep Space Nine}
\acro{ELTE}{Eotvos Lorand University}
\acro{ESA}{European Space Agency}
\acro{ESO}{European Space Organization}
\acro{ETC}{exposure time calculator}
\acro{FAQ}{Frequently Asked Questions}
\acro{FITS}{Flexible Image Transport System}
\acro{FOV}{field of view}
\acro{GEODSS}{Ground-Based Electro-Optical Deep Space Surveillance}
\acro{GIF}{Graphic Interchange Format}
\acro{GMOS}{Gemini Multi-Object Spectrograph}
\acro{GRFP}{Graduate Research Fellowship Program}
\acro{HARVEST}{Hunting for Activity in Repositories with Vetting-Enhanced Search Techniques}
\acro{HSC}{Hyper Suprime-Cam}
\acro{IAS15}{Integrator with Adaptive Step-size control, 15th order}
\acro{IAU}{International Astronomical Union}
\acro{IMACS}{Inamori-Magellan Areal Camera and Spectrograph}
\acro{IMB}{inner Main-belt}
\acro{IMCCE}{Institut de Mécanique Céleste et de Calcul des Éphémérides}
\acro{INT}{Isaac Newton Telescopes}
\acro{IP}{Internet Protocol}
\acro{IPAC}{Infrared Processing and Analysis Center}
\acro{IRSA}{Infrared Science Archive}
\acro{ITC}{integration time calculator}
\acro{JAXA}{Japan Aerospace Exploration Agency}
\acro{JD}{Julian Date}
\acro{JFC}{Jupiter Family Comet}
\acro{JPL}{Jet Propulsion Laboratory}
\acro{KBO}{Kuiper Belt object}
\acro{KIDS}{Kilo-Degree Survey}
\acro{KOA}{Keck Observatory Archive}
\acro{KPNO}{Kitt Peak National Observatory}
\acro{LBC}{Large Binocular Camera}
\acro{LBT}{Large Binocular Telescope}
\acro{LCOGT}{Las Cumbres Observatory Global Telescope}
\acro{LDT}{Lowell Discovery Telescope}
\acro{LINEAR}{Lincoln Near-Earth Asteroid Research}
\acro{LMI}{Large Monolithic Imager}
\acro{LONEOS}{Lowell Observatory Near-Earth-Object Search}
\acro{LSST}{Legacy Survey of Space and Time}
\acro{MBC}{Main-belt Comet}
\acro{MGIO}{Mount Graham International Observatory}
\acro{ML}{machine learning}
\acro{MMB}{middle Main-belt}
\acro{MOST}{Moving Object Search Tool}
\acro{MPC}{Minor Planet Center}
\acro{NASA}{National Aeronautics and Space Administration}
\acro{NAU}{Northern Arizona University}
\acro{NEA}{near-Earth asteroid}
\acro{NEAT}{Near-Earth Asteroid Tracking}
\acro{NEO}{near-Earth object}
\acro{NIHTS}{Near-Infrared High-Throughput Spectrograph}
\acro{NOAO}{National Optical Astronomy Observatory}
\acro{NOIRLab}{National Optical and Infrared Laboratory}
\acro{NONCOM}{Not Orbitally a Nominal Comet but Overtly a Minor planet}
\acro{NRC}{National Research Council}
\acro{OMB}{outer Main-belt}
\acro{OMBA}{outer main-belt asteroid}
\acro{OSIRIS-REx}{Origins, Spectral Interpretation, Resource Identification, Security, Regolith Explorer}
\acro{NSF}{National Science Foundation}
\acro{PANSTARRS}{Panoramic Survey Telescope and Rapid Response System}
\acro{Pan-STARRS}{Panoramic Survey Telescope and Rapid Response System}
\acro{PI}{Principal Investigator}
\acro{PNG}{Portable Network Graphics}
\acro{PSI}{Planetary Science Institute}
\acro{PSF}{point spread function}
\acro{PTF}{Palomar Transient Factory}
\acro{QHA}{Quasi-Hilda Asteroid}
\acro{QHC}{Quasi-Hilda Comet}
\acro{QHO}{Quasi-Hilda Object}
\acro{SAFARI}{Searching Asteroids For Activity Revealing Indicators}
\acro{SDSS}{Sloan Digital Sky Survey}
\acro{SMOKA}{Subaru Mitaka Okayama Kiso Archive}
\acro{SAO}{Smithsonian Astrophysical Observatory}
\acro{SBDB}{Small Body Database}
\acro{SDSS DR-9}{Sloan Digital Sky Survey Data Release Nine}
\acro{SLAC}{Stanford Linear Accelerator Center}
\acro{SOAR}{Southern Astrophysical Research Telescope}
\acro{SNR}{signal-to-noise ratio}
\acro{SSOIS}{Solar System Object Information Search}
\acro{SQL}{Structured Query Language}
\acro{STScI}{Space Telescope Science Institute}
\acro{SUP}{Suprime Cam}
\acro{SWRI}{Southwestern Research Institute}
\acro{TNO}{Trans-Neptunian object}
\acro{TRIF}{Technology and Research Initiative Fund} 
\acro{TSIP}{Telescope System Instrumentation Program}
\acro{UA}{University of Arizona}
\acro{UCSC}{University of California Santa Cruz}
\acro{UCLA}{University of California Los Angeles}
\acro{UCSF}{University of California San Francisco}
\acro{UT}{Universal Time}
\acro{VATT}{Vatican Advanced Technology Telescope}
\acro{VIRCam}{VISTA InfraRed Camera}
\acro{VISTA}{Visible and Infrared Survey Telescope for Astronomy}
\acro{VLT}{Very Large Telescope}
\acro{VST}{Very Large Telescope (VLT) Survey Telescope}
\acro{WFC}{Wide Field Camera}
\acro{WGSBN}{Working Group for Small Bodies Nomenclature}
\acro{WIRCam}{Wide-field Infrared Camera}
\acro{WISE}{Wide-field Infrared Survey Explorer}
\acro{WCS}{World Coordinate System}
\acro{YORP}{Yarkovsky--O'Keefe--Radzievskii--Paddack}
\acro{ZTF}{Zwicky Transient Facility}
\end{acronym}


\end{document}